\documentclass[%
reprint,
superscriptaddress,
showpacs,preprintnumbers,
amsmath,amssymb,
prb
]{revtex4-1}

\usepackage[pdftex]{graphicx}
\usepackage[pdftex]{color}
\usepackage{dcolumn}
\usepackage{bm}

\usepackage{color} 

\begin{document}

\preprint{APS/123-QED}

\title{Floquet-Weyl semimetals generated by an optically resonant interband-transition}

\author{Runnan Zhang}
\affiliation{Doctoral Program in Materials Science, Graduate School of Pure and Applied Sciences, University of Tsukuba, Tsukuba, Ibaraki 305-8573, Japan}
\author{Ken-ichi Hino}
\email{hino@ims.tsukuba.ac.jp}
\affiliation{Division of Materials Science, Faculty of Pure and Applied Sciences, University of Tsukuba, Tsukuba 305-8573, Japan}
\affiliation{Center for Computational Sciences, University of Tsukuba, Tsukuba 305-8577, Japan}
\author{Nobuya Maeshima}
\affiliation{Center for Computational Sciences, University of Tsukuba, Tsukuba 305-8577, Japan}
\affiliation{Division of Materials Science, Faculty of Pure and Applied Sciences, University of Tsukuba, Tsukuba 305-8573, Japan}

\date{\today}

\begin{abstract}

Floquet-Weyl semimetals (FWSMs) generated by irradiation of a continuous-wave laser with 
left-hand {\it circular} polarization (rotating in counterclockwise sense with time)
on the group II-V narrow gap semiconductor Zn$_3$As$_2$
are theoretically investigated, where the frequency of the laser is set nearly {\it resonant} with a band gap of the crystal.
It is found that the excitation of the crystal by such a laser induce two types of FWSM phases that differ absolutely in characters.
To be specific, the associated
two pairs of Weyl points are stably formed by band touching between Floquet sidebands ascribable to a valence band labeled as $J_z=\pm3/2$ and a conduction band labeled as $J_z=\pm 1/2$,
where $J_z$ represents the $z$-component of total angular momentum quantum number of $\Gamma$-point and a double sign corresponds.
Here, one FWSM state composed of the up-spin Floquet sidebands relevant to $J_z=3/2$ and $1/2$ shows almost quadratic band-touching in the vicinity of the associated pair of Weyl points, while the other FWSM state composed of the down-spin Floquet sidebands relevant to $J_z=-3/2$ and $-1/2$ shows linear band-touching.
Further, it is revealed that both up-spin and down-spin sidebands host
nontrivial two-dimensional surface states that are pinned to the respective 
pairs of the Weyl points.
Both surface states also show different energy dispersions and physical properties. 
More detailed discussion is made in the text on the origin of the above findings, chirality of the FWSM phases, alteration of topological order, laser-induced magnetic properties, and so on.

\end{abstract}

\maketitle


\section{Introduction}
\label{sec1}

Topological materials have been studied for more than a decade 
\cite{Kane2005,Bernevig2006,Hasan2010,Qi2011}
and growing interest has been directed toward the exploration of a class of topological semimetals (SMs) 
\cite{Murakami2007a,Murakami2007b,Wang2012,Wang2013,Young2012,Young2015,Young2015,
Yang2014,Yi2014,Xu2015,Kargarian2016,Park2017,Doh2017,Yan2017,Armitage2018,Yang2018,Ramankutty2018,
Luo2020}
--- such as Weyl SMs (WSMs), Dirac SMs (DSMs), and nodal-line SMs (NLSMs) --- in addition with further deepening of the studies of prototypical topological insulators, for instance, on non-Hermitian topological systems,\cite{Kunst2018,Yao2018}
topological photonics,\cite{Ozawa2019}
higher-order topological insulators,
\cite{Benalcazar2017,Langbehn2017,Peng2017}
topological excitons,
\cite{Budich2014,Entina2016,Chen2017}
topological surface plasmon polaritons,
\cite{Li2014,Deshko2016}
and so on.
WSMs and DSMs are three dimensional (3D) gapless phases of materials, in which 
bands cross linearly at points
protected by topology and symmetry.
There is close connection with the 
chiral anomaly (the Adler-Bell-Jackiw anomaly) in linearly dispersing fermionic
excitations in particle physics, which gives rise to the nonconservation of 
an axial current even for a massless particle.
\cite{Weyl1929,Majorana1937,Gynther2011,Elliott2015,Adler1969,Bell1969,Nielsen1983}
In accord with the prediction of the chiral anomaly, large negative magnetic resistance is observed in magnetotransport in WSMs. 
\cite{Son2013,Burkov2013,Burkov2014,Xiong2015}
Further, these SMs exhibit a great number of novel transport properties such as 
ultra-high mobility, titanic magnetic resistance, and anomalous Hall conductivity.
\cite{Wang2013,Jeon2014,Feng2015,Liang2015,Wang2016,Son2013,Arnold2016,Huang2015,
Ali2014,Shekhar2015,Yan2017}
In NLSMs, bands cross along special lines in the Brillouin zone (BZ) in the shape of 
a closed ring or a line.
Breaking of either time- reversal (T-) symmetry or spatial-inversion (I-) symmetry leads DSMs and NLSMs to WSMs.\cite{Yang2018}
In the T-breaking WSMs, there are a pair of Weyl nodes with opposite chirality,
on which a surface state is pinned with a characteristic Fermi arc, while
the number of Weyl nodes in the I-breaking WSMs is a multiple of four.
\cite{Murakami2007b,Armitage2018}

An interaction of topological SMs with a continuous-wave laser provides the studies of
topological materials with another avenue from the perspective of the quantum control of
underlying topological properties by means of built-in laser parameters --- intensity, frequency
$\omega$, and polarization --- and the exploration of topological phases
that are in non-equilibrium.
\cite{Wang2014,Taguchi2016a,Taguchi2016b,Hubener2017,Zou2016,Zhang2016,Chen2018,
Kumar2019,Zhu2020,Salerno2020,Gao2020,Kawaguchi2020,
Tokman2020,Zhang2021,Liang2021,Ma2015,Juan2017,Ghorashi2018,Umer2021a,Umer2021b}
Here, the total Hamiltonian $H(t)$ of concern at time $t$ has temporal periodicity 
$H(t)=H(t+T)$ to ensure the Floquet theorem with $T=2\pi/\omega$.
\cite{Shirley1965}
By the drive with a circularly polarized laser --- in place of the application of a static intrinsic Zeeman field ---,
the T-symmetry in DSMs and NLSMs is broken to form WSMs, and these are termed as Floquet WSMs (FWSMs).
This scenario for creating FWSMs is applied to the DSMs of alkali pnictides NaBi$_3$,\cite{Hubener2017} type-II and hybrid LNSMs,\cite{Chen2018}
and
3D stacked graphene systems.\cite{Zou2016}
In particular, it is reported that the drive of a 3D Luttinger semimetal by 
an elliptically polarized light leads to the coexistence of WSM phases with 
double and single Weyl points, which can be tuned to be type-I or type-II.
\cite{Ghorashi2018}
Further, by virtue of Floquet engineering due to periodical driving of 
hybrid multi-WSM phases, the number of various isolated band touching points
can be increased on-demand by tuning system parameters, where intricate Fermi arc structures are hosted.
\cite{Umer2021a}
In addition, NLSMs are driven to result in FWSMs, revealing 
a photovoltaic anomalous Hall effect 
associated with the Weyl point nodes.\cite{Taguchi2016b}
Very recently, frequency-independent magnetization mechanisms in response to
circularly polarized light are studied in WSMs.
\cite{Gao2020,Tokman2020}

As regards the I-symmetry, this is also broken by the introduction of an interaction of electron with the continuous-wave laser.
However, the time-glide I-symmetry holds correct instead to realize 
the same invariance in $H(t)$ as the I-symmetry.\cite{Zhang2021,Morimoto2017}
The symmetry operation associated with this symmetry is represented by
the operation of putting time $t$ ahead by a half period $T/2$, followed by the I-operation.

In most of theoretical studies of Floquet topological materials, 
\cite{Kitagawa2010,Oka2009,Zhenghao2011,Lindner2011,Rudner2013,Rechtsman2013,Claassen2017,
Kitamura2017,Hansen2017,Nakagawa2020}
an electron-light interaction is introduced exclusively by employing the Peierls phase transform
--- given by the replacement of a Bloch momentum $\boldsymbol{k}$ by $\boldsymbol{k}+\boldsymbol{A}(t)$ --- under the off-resonant condition that
$\omega \gg E_g$.
Here, $\boldsymbol{A}(t)$ and $E_g$ represent a vector potential of the laser at $t$, and 
a bandgap of the concerned material, respectively, and the atomic units are used.
Further, the effective Floquet Hamiltonian is constructed by relying on the Floquet-Magnus expansion with respect to $E_g/\omega$.\cite{Mananga2011,Mananga2016,Haga2019}
It is remarked that the convergence of this expansion is not ensured
at the resonant limit of $\omega=E_g$.\cite{Haga2019}
The resulting Floquet bands are likely modified from original bands before laser irradiation.
It is assumed that under the above off-resonant condition, effects of interband electric-dipole transitions between a valence band and a conduction band
are negligibly small,
compared with those of the above-mentioned Peierls interaction. \cite{Wang2014,Hubener2017,Hansen2017}
Here, interband and intraband couplings caused by the electron-light interaction due to the
Peierls phase transform are termed as the Peierls interaction to make a distinction from the interband electric-dipole interaction.

The aim of this paper is to create a FWSM phase by driving Zn$_3$As$_2$
\cite{LinChung1969,Okamoto1992,Sieranski1994,Botha1999}
belonging to the group II-V semiconductors 
with a circularly polarized laser which meets an almost on-resonant condition $\omega \approx E_g$,\cite{Lindner2011,Zhang2021} and to explore the properties of surface states hosted by this FWSM.
It is remarked that this material is topologically trivial without band inversion.
This differs from a DSM of Cd$_3$As$_2$,\cite{LinChung1969,Sieranski1994,Neupane2014,Liu2014,Crassee2018} realized 
by the band-inversion mechanism due to 
the presence of a $n(>2)$-fold uniaxial rotational
symmetry along a symmetry line,
hosting edge modes known as double Fermi arcs at the surfaces.
\cite{Wang2012,Wang2013,Yang2014,Kargarian2016,Armitage2018}
The creation of the FWSM is governed by an almost resonant transition due to the
interband electric-dipole interaction rather than that due to the Peierls interaction, as shown in more detail later (Sec.~\ref{sec3}).
This is the key issue of this paper.

Below, a more detailed explanation is made on this key issue
based on the four-band model of the semiconductor Zn$_3$As$_2$, where
the valence and conduction bands are labeled as $J_z=\pm3/2$ and $\pm 1/2$,
respectively, with $J_z$ as the $z$-component of total angular momentum quantum number at the $\Gamma$-point ($\boldsymbol{k}=\boldsymbol{0}$).
First, given the driving laser with a left-hand circular polarization (rotating in 
counterclockwise sense with time), the coupling of this light with 
an electric dipole moment induced by the transition between the down-spin bands with $J_z=-3/2$ and $-1/2$ is dominant over that between the up-spin bands 
with $J_z=3/2$ and $1/2$.\cite{Allen2003}
This is maximized when the on-resonant condition is met.
It is remarked that the roles of the up-spin and down-spin bands are exchanged
for a laser with a right-hand circular polarization.
A left-hand polarization is favored throughout this paper unless otherwise stated.
Second, as the laser intensity increases, the ac-Stark effect gives rise to
larger energy splitting of each down-spin band into two leaves
with conspicuous modification of the band profile,
\cite{Zhang2021,Autler1955,Knight1980,Sie2015}
whereas the up-spin bands are just slightly affected.
The ac-Stark effect is also maximized by the on-resonant condition.
Third, the present resonant interband-transition yields real carrier excitation, differing from virtual carrier excitation due to the off-resonant one.
Thus, it is likely that orbital magnetization results from the inverse Faraday effect
that is a non-linear optical process caused by a circularly polarized laser field.
\cite{Gao2020,Kawaguchi2020,Tokman2020,Liang2021,Pershan1966,Kimel2005,Hertel2006,
Zhang2009,Battiato2014}
Consequently, it is expected that the laser drive with a circular polarization in the almost on-resonant condition provides intriguing physics with FWSMs, which is sharply distinct from the conventional 
off-resonant laser drive.\cite{Wang2014,Hubener2017,Hansen2017}

The remainder of this paper is organized as follows. 
Section~\ref{sec2} describes the theoretical framework,
Sec.~\ref{sec3} presents the results and discussion, and Sec.~\ref{sec4} presents the conclusion. 
Further, three appendices are included.
Hereafter,
the atomic units (a.u.) are used throughout, unless otherwise stated.

\section{Theory}
\label{sec2}

\subsection{Effective Hamiltonian}
\label{sec2A}

The crystal of concern, Zn$_3$As$_2$, is a narrow gap semiconductor,
the structure of which is very similar to that of Cd$_3$As$_2$,
though in the latter, a band is inverted to result in a DSM.
\cite{LinChung1969,Sieranski1994,Neupane2014,Liu2014,Crassee2018}
There are many equilibrium phases of Zn$_3$As$_2$ depending on pressure and temperature,
for instance, $\alpha$Zn$_3$As$_2$ (with a body-centered tetragonal structure I4$_1/$cd)
and $\alpha^\prime$Zn$_3$As$_2$ (with a tetragonal structure P4$_2/$nbc).
\cite{Okamoto1992}
Here, one employs the structure of $\alpha^\prime$Zn$_3$As$_2$
having the C$_4$-rotational symmetry along the $\Gamma-$Z axis in the BZ
for constructing an effective Hamiltonian.
The low-energy electronic
properties of it are mostly determined by the conduction band composed of Zn 4$s$-orbitals
and the valence band composed of As 4$p$-orbitals.

Here, an effective electronic Hamiltonian for Zn$_3$As$_2$ is constructed by
following the Kane model used in Ref.~\onlinecite{Wang2013} for the crystal structure of
Cd$_3$As$_2$ with a tetragonal structure P4$_2/$nbc.
To be specific, one considers 
the following four states as conduction $s$-states
$|\Gamma_6, J_z=\pm 1/2\rangle$ and heavy-hole $p$-states $|\Gamma_7, J_z=\pm 3/2\rangle$, and light-hole states and split-off states are disregarded 
because of relatively large energy 
separation from these four states at the $\Gamma$-point.
The effective Hamiltonian is read as the $4 \times 4$-matrix:
\cite{Kane1957,Luttinger1956}
\begin{equation}
\mathcal{H}(\boldsymbol{k})
=c(\boldsymbol{k})I+\sum_{j=3}^5 d_j(\boldsymbol{k})\gamma_j
\label{calH}
\end{equation}
with $\boldsymbol{k}=(k_x, k_y,k_z)$ as a 3D Bloch momentum.
Here, $\gamma_j$'s represent the
four-dimensional Dirac matrices for the Clifford algebra, defined by
\(
\gamma_1=\tau_x\otimes\sigma_x,\:
\gamma_2=\tau_x\otimes\sigma_y,\:
\gamma_3=\tau_x\otimes\sigma_z,\:
\gamma_4=\tau_z\otimes I_2
\), and 
\(
\gamma_5=\tau_y\otimes I_2
\), where
$I$ and $I_2$ represent the $4\times 4$ and $2\times 2$ unit matrices, respectively, 
$\tau_l$ and $\sigma_l$ with $l=x,\:y,\:z$ represent
the Pauli matrices for orbital and spin degrees of freedom, respectively, and 
the anti-commutation relation,
\(
\{\gamma_j, \gamma_{j^\prime} \}=2\delta_{jj^\prime}
\), 
is ensured.
According to the above definition of $\gamma_j$'s, it is understood that
the states of
$|\Gamma_6, J_z=1/2\rangle,\: |\Gamma_7, J_z= 3/2\rangle,\:
|\Gamma_6, J_z=-1/2\rangle$,
and
$|\Gamma_7, J_z=-3/2\rangle
$
are labeled as $1, 2, 3$, and 4, respectively,
for the matrix elements of
$\mathcal{H}(\boldsymbol{k})$, namely, $\{\mathcal{H}_{mn}(\boldsymbol{k})\}$ 
with $m,n=1\sim4$.
Moreover, $d_j(\boldsymbol{k})$'s are given by
\begin{eqnarray}
\left\{
\begin{array}{l}
d_3(\boldsymbol{k})=t_{sp}\sin{(k_xd_x)}\\
d_4(\boldsymbol{k})=\Delta_g
+\sum_{l=x,y,z}\epsilon_l(k_l)\\
d_5(\boldsymbol{k})=t_{sp}\sin{(k_yd_y)}
\end{array}
\right.,
\label{dk}
\end{eqnarray}
where
\begin{equation}
\epsilon_l(k_l)=-2t^{xy}\{1-\cos{(k_ld_l)}\}
\label{epsilonl0}
\end{equation}
for $l=x,y$,
and
\begin{equation}
\epsilon_z(k_z)=-2t^z\{1-\cos{(k_zd_z)}\}.
\label{epsilonz0}
\end{equation}
Here, $t^l$ represents a hopping matrix between identical bands in the $l$-direction
with $l=x,y,z$, where $t^l<0$, $t^{xy}\equiv t^x=t^y$, and
$t_{sp}$ represents a hopping matrix between different bands due to a spin-orbit coupling.
Further, $d_l$ represents a lattice constant in the $l$-direction, and
the bandgap at the $\Gamma$-point $E_g$ is given by $E_g=E_{\Gamma_6}-E_{\Gamma_7}$ with $\Delta_g=E_g/2$, where the conduction and valence band energies at $\Gamma_6$ and $\Gamma_7$ 
are represented as $E_{\Gamma_6}$ and $E_{\Gamma_7}$, respectively.
An additional energy $c(\boldsymbol{k})$ is given by
\begin{equation}
c(\boldsymbol{k})=E_F
+2\sum_{l=x,y,z}m^{l}\{1-\cos{(k_ld_l)}\}
\end{equation}
with $m^l$'s constants,
and the Fermi energy $E_F$ is set equal to zero: $E_F=0$.
Hence, 
$E_{\Gamma_6}=c(\boldsymbol{0})+d_4(\boldsymbol{0})$ and
$E_{\Gamma_7}=c(\boldsymbol{0})-d_4(\boldsymbol{0})$.

It is assumed that the off-diagonal block matrices of $\mathcal{H}(\boldsymbol{k})$ have
little contributions to the band structure under the present tetragonal symmetry,
that is, $d_1(\boldsymbol{k}), d_2(\boldsymbol{k}) \approx0$, leading to
$[I_2 \otimes \sigma_z,\mathcal{H}(\boldsymbol{k})]\approx 0$.\cite{comment2}
Thus, $\mathcal{H}(\boldsymbol{k})$ is cast into the block-diagonal form
\begin{equation}
\mathcal{H}(\boldsymbol{k})
=\left(
\begin{array}{cc}
h(\boldsymbol{k})& 0\\
0& h^*(-\boldsymbol{k})
\end{array}
\right),
\label{Hcal2}
\end{equation}
where
\(
h(\boldsymbol{k})=d_3(\boldsymbol{k})\tau_x+d_4(\boldsymbol{k})\tau_z
+d_5(\boldsymbol{k})\tau_y.
\)

An interaction of electron with light is introduced into $\mathcal{H}(\boldsymbol{k})$
by replacing $\boldsymbol{k}$ by $\boldsymbol{K}(t)=\boldsymbol{k}+\boldsymbol{A}(t)$, followed by adding to $\mathcal{H}(\boldsymbol{K}(t))$ an interband electric dipole
interaction represented by
\(
\mathcal{H}^\prime(t).
\)
\cite{comment1}
Here, the replacement by $\boldsymbol{K}(t)$ results from the Peierls phase transform 
in the lattice representation of the effective Hamiltonian, 
and the optical interaction arising from this replacement is herein termed as the Peierls interaction, as mentioned in Sec.~\ref{sec1}.
Further, the interband electric-dipole interaction is provided as 
$\mathcal{H}^\prime(t)=\boldsymbol{F}(t)\cdot\boldsymbol{M}$, where
$\boldsymbol{M}$ represents a matrix of electric-dipole transition between
$|\Gamma_6, J_z=\pm 1/2\rangle$ and $|\Gamma_7, J_z=\pm 3/2\rangle$, independent of $\boldsymbol{k}$; a double sign corresponds.
The vector potential is given by
\begin{equation}
\boldsymbol{A}(t)=\left(
-{F_x\over \omega} \sin\omega t,{F_y\over \omega} \cos\omega t,0
\right)
\label{A}
\end{equation}
with $F_x$ and $F_y$ constants,
and in view of $\boldsymbol{F}(t)=-\dot{\boldsymbol{A}}(t)$,
the associated electric field becomes
\begin{equation}
\boldsymbol{F}(t)=\left(
F_x\cos\omega t,F_y\sin\omega t,0
\right).
\label{F}
\end{equation}
The laser is linearly polarized in the $x$-direction when $F_x\not=0$ and 
$F_y=0$,
while left-hand circularly polarized in the $x-y$ plane
when $F_x$ is set equal to $F_y$, namely, 
$F_c\equiv F_x=F_y$.
The time-dependent effective Hamiltonian of the driven semiconductor is thus read as
\cite{comment1}
\begin{equation}
H(\boldsymbol{k},t)=C(\boldsymbol{k},t)I
+\sum_{j=3}^5D_j(\boldsymbol{k},t)\gamma_j
+\mathcal{H}^\prime(t),
\label{H}
\end{equation}
where
$C(\boldsymbol{k},t)\equiv c(\boldsymbol{K}(t))$ and 
$D_j(\boldsymbol{k},t)\equiv d_j(\boldsymbol{K}(t))$.

Obviously, this Hamiltonian ensures the temporal periodicity, $H(\boldsymbol{k},t+T)=H(\boldsymbol{k},t)$
with $T=2\pi/\omega$, and the system of concern follows the Floquet theorem.\cite{Shirley1965}
Thus, the present time-dependent problem ends up with the following Floquet eigenvalue problem as:
\begin{equation}
L(\boldsymbol{k},t)\Psi_\alpha(t)=E_\alpha(\boldsymbol{k})\Psi_\alpha(t),
\label{Floquet}
\end{equation}
where
\begin{equation}
L(\boldsymbol{k},t)=H(\boldsymbol{k},t)-iI{\partial\over \partial t},
\label{L}
\end{equation}
$E_\alpha(\boldsymbol{k})$ represents the $\alpha$th eigenvalue termed a quasienergy
or a Floquet energy, and $\Psi_\alpha(t)$ represents the associated eigenvector ensuring
the temporal periodicity,
$\Psi_\alpha(t+T)=\Psi_\alpha(t)$.
In actual calculations, a set of $E_\alpha(\boldsymbol{k})$'s are 
obtained by numerically solving Eq.~(\ref{Floquet}) in the $\omega$-domain,
where the Floquet matrix $L(\boldsymbol{k},t)$ is recast into 
a Fourier-Floquet matrix element
$\tilde{L}_{nn^\prime}(\boldsymbol{k},\omega)$
with respect to $n$ and $n^\prime$ photon states.\cite{Shirley1965}
This is read as
\begin{eqnarray}
\tilde{L}_{nn^\prime}(\boldsymbol{k},\omega)
&=&
\tilde{C}_{nn^\prime}(\boldsymbol{k},\omega)I
+\sum_{j=3}^5 \tilde{D}_{j,nn^\prime}(\boldsymbol{k},\omega)\gamma_j
\nonumber\\
&&
+\tilde{\mathcal{H}^\prime}_{nn^\prime}(\omega)
+n\omega I\delta_{nn^\prime},
\label{Ltilde}
\end{eqnarray}
where it is understood that the Fourier transform of matrix $X(t)$ is represented by
\begin{equation}
\tilde{X}_{nn^\prime}(\omega)=
{1\over T}\int^T_0 dt\:e^{-i(n-n^\prime)\omega t}X(t).
\end{equation}
In addition, it is remarked that in fact, $\tilde{C}_{nn^\prime}(\boldsymbol{k},\omega)$
is less dependent on the set of photon numbers, $n$ and $n^\prime$, and 
almost identical to $c(\boldsymbol{k})$.
Hence, hereafter, it is understood that $E_\alpha(\boldsymbol{k})$'s are reckoned from
$c(\boldsymbol{k})$; in other words, the effect of $c(\boldsymbol{k})$
on $E_\alpha(\boldsymbol{k})$'s is neglected.
Explicit expressions of $\tilde{D}_{j,nn^\prime}(\boldsymbol{k},\omega)\;\;(j=3\sim5)$
are given in Appendix~\ref{app1}.

\subsection{Electric dipole couplings}
\label{sec2B}

It is convenient to describe an explicit form of $\mathcal{H}^\prime(t)$.
This is provided as
\begin{equation}
\mathcal{H}^\prime(t)
=(\Omega_y\sin{\omega t})\tau_x\otimes I_2
+(\Omega_x\cos{\omega t}) \tau_y\otimes \sigma_z,
\label{Hprime}
\end{equation}
where
$\Omega_x=F_xP/\sqrt{2}$ and $\Omega_y=F_yP/\sqrt{2}$.
Here, $P$ is a dipole matrix element given by
$P=\langle S|x|X\rangle =\langle S|y|Y\rangle$,
where
$x$ and $y$ represent the $x$ and $y$ components of electron position $\boldsymbol{r}$,
respectively,
and the states of
$|\Gamma_6, J_z=\pm 1/2\rangle$ and $|\Gamma_7, J_z=\pm 3/2\rangle$ are
represented by
$|\Gamma_6, J_z=\pm 1/2\rangle=i|S\rangle$
and
$|\Gamma_7, J_z=\pm 3/2\rangle=\pm(1/\sqrt{2})|X\pm i Y\rangle$,
respectively,
in terms of $s, p_x$, and $p_y$ states denoted as
$|S\rangle, |X\rangle$, and $|Y\rangle$, respectively.
Similar to Eq.~(\ref{Hcal2}), $\mathcal{H}^\prime(t)$ is block-diagonalized 
as
\begin{equation}
\mathcal{H}^\prime(t)
=\left(
\begin{array}{cc}
V^{(+)}(t)& 0\\
0& V^{(-)}(t)
\end{array}
\right),
\label{V}
\end{equation}
where
$V^{(+)}(t)$ and $V^{(-)}(t)$ represent the electric dipole couplings
between the up-spin bands,
$|\Gamma_6, J_z=1/2\rangle$ and $|\Gamma_7, J_z= 3/2\rangle$, 
and the down-spin bands, $|\Gamma_6, J_z=-1/2\rangle$ and $|\Gamma_7, J_z= -3/2\rangle$,
respectively, given by
\begin{equation}
V^{(\pm)}(t)
=(\Omega_y\sin{\omega t}) \tau_x
\pm(\Omega_x\cos{\omega t}) \tau_y,
\label{V+-}
\end{equation}
where a double sigh corresponds.
The Fourier transform of $V^{(\pm)}(t)$ into the $\omega$-domein is given in Appendic~\ref{app1}.
In view of Eqs.~(\ref{H}) and (\ref{V}),
$H(\boldsymbol{k},t)$ is cast into the block-diagonalized form:
\begin{equation}
H(\boldsymbol{k},t)
=C(\boldsymbol{k},t)I
+\left(
\begin{array}{cc}
H^{(+)}(\boldsymbol{k},t)& 0\\
0& H^{(-)}(\boldsymbol{k},t)
\end{array}
\right),
\label{H2}
\end{equation}
that is,
$[I_2 \otimes \sigma_z,H(\boldsymbol{k},t)]=0$.
Here, the up-spin Hamiltonian $H^{(+)}(\boldsymbol{k},t)$ and
the down-spin Hamiltonian $H^{(-)}(\boldsymbol{k},t)$ are given by
\begin{eqnarray}
H^{(\pm)}(\boldsymbol{k},t)&=&
\left[
\pm D_3(\boldsymbol{k},t)+\Omega_y\sin{\omega t}
\right]\tau_x
+D_4(\boldsymbol{k},t)\tau_z \nonumber\\
&&
+
\left[
D_5(\boldsymbol{k},t)\pm\Omega_x\cos{\omega t}
\right]\tau_y.
\label{H+-}
\end{eqnarray}

The expression of Eq.~(\ref{V+-}) implies that in general, an optical dipole interaction between up-spin
bands
is different from that between down-spin
bands.
To be specific, for a linearly polarized light,
\begin{equation}
V^{(\pm)}(t)=
\pm(\Omega_x\cos{\omega t})\tau_y
=
\pm\Omega_x\cos{\omega t}
\left(
\begin{array}{cc}
0&-i\\
i&0
\end{array}
\right)
\label{Vlin}
\end{equation}
with $\Omega_y=0$,
and $V^{(+)}(t)$ is identical to $V^{(-)}(t)$
aside from an unimportant phase factor $\mp i$.
On the other hand,
for a left-hand circularly polarized light,
\begin{eqnarray}
V^{(\pm)}(t)&=&
\Omega_c
\left[
(\sin{\omega t})\tau_x
\pm
(\cos{\omega t})\tau_y
\right]
\nonumber\\
&=&
\mp i\Omega_c
\left(
\begin{array}{cc}
0&e^{\pm i\omega t}\\
-e^{\mp i\omega t}&0
\end{array}
\right)
\label{Vcir}
\end{eqnarray}
with $\Omega_c\equiv\Omega_x=\Omega_y$,
and $V^{(+)}(t)$ and $V^{(-)}(t)$ are different from each other.
In particular, this distinction stands out for a linear optical transition,
for instance, from the valence band at $\Gamma_7$ to 
the conduction band at $\Gamma_6$.
In view of the matrix element $V^{(\pm)}_{12}(t)$ of Eq.~(\ref{Vcir}),
the transition amplitudes of the photoabsorption between the up-spin bands, represented as
$a^{(+)}$, and that between the down-spin bands, represented as
$a^{(-)}$, 
are given by
\begin{equation}
a^{(\pm)}=\mp i\Omega_c\int^\infty_{-\infty} dt\: e^{i(E_{\Gamma_6}-E_{\Gamma_7}\pm\omega)t}
\propto \delta(E_g\pm\omega)
\end{equation}
with $E_g=E_{\Gamma_6}-E_{\Gamma_7}>0$.
It is evident that 
the transition between the up-spin bands is forbidden, namely, $a^{(+)}=0$,
while that between the down-spin
bands is allowed, namely, $a^{(-)}\not=0$,
because of the energy conservation $\omega=E_g$.
Incidentally, as regards the related photoemission, 
in view of the matrix element $V^{(\pm)}_{21}(t)$,
the transition amplitudes of it are given by
\begin{equation}
b^{(\pm)}=\pm i\Omega_c\int^\infty_{-\infty} dt\: e^{i(E_{\Gamma_7}-E_{\Gamma_6}\mp\omega)t}
\propto \delta(E_g\pm\omega),
\end{equation}
and the same discussion as the photoabsorption is applicable;
$b^{(+)}=0$ and $b^{(-)}\not=0$.

As long as $\omega \approx E_g$,
these results almost hold correct in non-linear optical processes
including strongly photoinduced processes,
though the contribution from up-spin bands does not vanish because
the energy conservation is not required in virtual states.
In other words, the effect of $V^{(-)}(t)$ is dominant over that of
$V^{(+)}(t)$.
This is one of the key issues in this paper, as mentioned in Sec.~\ref{sec1}.
In contrast, as regards off-resonant cases that $\omega \gg E_g$ or $\omega \ll E_g$,
$V^{(+)}(t)$ and $V^{(-)}(t)$ would have almost equal, however, vanishingly small contributions 
to optical processes, as long as 
$ |E_g-\omega|\gg 2\Omega_c$;
for more detail, see Eq.~(\ref{bandsn}) or (\ref{bandpn}) to be shown later.

\subsection{Symmetries}
\label{sec2C}

It is evident that 
both T- and I-symmetries are conserved in $\mathcal{H}(\boldsymbol{k})$,
that is,
\(
\Theta^{-1}\: \mathcal{H}(-\boldsymbol{k}) \Theta=\mathcal{H}(\boldsymbol{k})
\), and
\(
\Pi^{-1}\: \mathcal{H}(-\boldsymbol{k}) \Pi=\mathcal{H}(\boldsymbol{k})
\), 
where $\Theta$ and $\Pi$ represent the T- and I-operators, defined by
$\Theta=-iI_2\otimes \sigma_y K$ and $\Pi=\tau_z\otimes I_2$, respectively, where 
$K$ means an operation of taking complex conjugate.
Further,
the T-symmetry is still respected in $H(\boldsymbol{k},t)$ for a linearly polarized light,
that is,
\(
\Theta^{-1} H(-\boldsymbol{k},-t) \Theta=H(\boldsymbol{k}, t)
\),
and thus, a pair of up-spin and down-spin Floquet bands forms Kramers degeneracy.
On the other hand,
the T-symmetry is broken for a circularly polarized light,
that is,
\(
\Theta^{-1} H(-\boldsymbol{k},-t) \Theta\not=H(\boldsymbol{k}, t)
\).

As regards the I-symmetry, this is broken, 
that is, 
\(
\Pi^{-1} H(-\boldsymbol{k},t) \Pi\not =H(\boldsymbol{k},t)
\),
because 
$D_j(-\boldsymbol{k},t) \not=-D_j(\boldsymbol{k},t)$ for $j=$3 and 5, $D_4(-\boldsymbol{k},t) \not=D_4(\boldsymbol{k},t)$,
and
$\Pi^{-1} \mathcal{H}^\prime(t) \Pi\not =\mathcal{H}^\prime(t)$.
In fact, it is shown that in terms of an operator defined as 
$\tilde{\Pi}=\Pi \mathcal{T}_{1/2}$, the symmetry
\(
\tilde{\Pi}^{-1} H(-\boldsymbol{k},t+T/2) \tilde{\Pi}=H(\boldsymbol{k},t)
\)
is retrieved,
where $\mathcal{T}_{1/2}$ represents the operation of putting $t$ ahead by a half period $T/2$,
namely, the replacement of $t \rightarrow t+T/2$.\cite{Zhang2021,Morimoto2017}
This is the time-glide $I$-operator mentioned in Sec.~\ref{sec1}.
Therefore, despite the breaking of the I-symmetry, 
a Floquet band disperses in a symmetric manner with respect to $\boldsymbol{k}$, namely, $E_\alpha(\boldsymbol{k})=E_\alpha(-\boldsymbol{k})$.
For a linearly polarized light,
it is still probable that a four-fold band crossing occurs at the high-symmetry points,
namely, the time-reversal invariant momenta.

\section{Results and Discussion}
\label{sec3}

In the actual calculations, the following material parameters\cite{Wang2012,Wang2013} and laser parameters are employed as:
$E_g=0.0169\: (0.46 \:\text{eV}), \omega=0.0147\: (0.4 \:\text{eV}), d_x=d_y=5.67\: (3\text{\AA}), d_z=9.44\: (5\text{\AA}), t^{xy}=-0.0018, t^z=-0.0074, t_{sp}=0.0037,
F_x=F_c=0.0003\:(1.54\:\text{MV/cm})$, and $P=25.9$.
Further, the maximum
number of photons $(N_p)$ incorporated in the calculations is set to be three to reach numerical convergence, that is, $n,n^\prime=-N_p\sim N_p$ for the Fourier-Floquet matrix $\tilde{L}_{nn^\prime}(\boldsymbol{k},\omega)$.

Based on these numerical parameters, one evaluates the degree of magnitude of effects due to the Peierls interaction.
These contributions are determined in terms of factors given by
the $n$th-order Bessel function of the first kind $J_n(z_l)$ that is included in
$\tilde{D}_{j,nn^\prime}(\boldsymbol{k},\omega)\;(j=3\sim 5)$ of
Eqs.~({\ref{appD3})-(\ref{appD5}) with 
$z_l \equiv F_ld_l/\omega\;(l=x,y)$.
For $z_l=0.116$, one obtains that $J_0(z_l)-1=-0.0034, J_1(z_l)=0.056$, and $J_2(z_l)=0.0017$, implying that the hopping matrices of $t^l$ and $t_{sp}$ are modified
just by the order of $10^{-2}\sim 10^{-3}$ by the Peierls interaction.
Thus, it is stated that in the system of concern, this interaction plays a less significant role 
than the interband electric-dipole interaction.

\subsection{Qualitative Understanding of Band Structures}
\label{sec3A}

It is preferable to show an overall Floquet band structure in the present system
in a qualitative manner prior to the discussion of rather complicated numerical results.
Here, a Floquet state $\alpha$ attributed to a $b$-band dressed with $n$ photons is 
denoted as $b(n)$ with $b=e,hh$, 
where the bands $e$ and $hh$ represent the $s$ and heavy-hole $p$ orbitals, respectively.
Below, one seeks approximate Floquet bands represented 
in an analytic closed form in a two-step manner.
First, based on a two-band model incorporating $e(n-1)$ and $hh(n)$ for each spin state, 
one seeks expressions of hybridized bands of states $e(n-1)$ 
for up- and down-spins, represented as
$\mathcal{E}^{(+)}_{e(n-1)}$ and $\mathcal{E}^{(-)}_{e(n-1)}$,
respectively.
Here, the ac-Stark effect with a Rabi frequency $\Omega_l$ is incorporated by employing the 
rotational-wave approximation.
Similarly, the coupling between $e(n)$ and $hh(n+1)$ for each spin state 
results in expressions of hybridized bands of states $hh(n+1)$ 
for up- and down-spins, represented as
$\mathcal{E}^{(+)}_{hh(n+1)}$ and $\mathcal{E}^{(-)}_{hh(n+1)}$,
respectively.
Second, by introducing the residual spin-orbit interaction between
$\mathcal{E}^{(\pm)}_{e(n-1)}$ and $\mathcal{E}^{(\pm)}_{hh(n+1)}$ bands,
one obtains the desired expressions of hybridized Floquet bands represented by
$E^{(\pm)}_{e(n-1)}(\boldsymbol{k})$ and $E^{(\pm)}_{hh(n+1)}(\boldsymbol{k})$: 
a double sign corresponds.
For more detail of the derivation, consult Appendices~\ref{app2-1}-\ref{app2-3}.

In the case that the band $hh(n+1)$ is located above the band $e(h-1)$,
the resulting energy, represented by $E^{(\pm)}(\boldsymbol{k})$, 
is cast into
$E^{(\pm)}(\boldsymbol{k})=E^{(\pm)}_{hh(n+1)}(\boldsymbol{k}) \ge n\omega$
and $E^{(\pm)}(\boldsymbol{k})=E^{(\pm)}_{e(n-1)}(\boldsymbol{k}) \le n\omega$,
the expressions of which are
given by Eqs.~(\ref{Ebeta}) and (\ref{E0beta}), respectively.
To avoid unnecessary complication in these expressions,
the approximations that
$J_0(z_l)\approx 1$ and $J_{n(\not=0)}(z_l)\approx 0$ are made.
Thus, these are read as
\begin{eqnarray}
E^{(\pm)}_{e(n-1)}(\boldsymbol{k})
&\approx&-\left[\left\{
\omega/2
-\left(\eta(\boldsymbol{k})^2+|\mathcal{W}_0^{(\pm)}|^2
\right)^{1/2}
\right\}^2
\right.\nonumber\\
&&
\left.
+|\mathcal{V}_0^{(\pm)}|^2
\right]^{1/2}+n\omega,
\label{bandsn}
\end{eqnarray}
and
\begin{eqnarray}
E^{(\pm)}_{hh(n+1)}(\boldsymbol{k})
&\approx&\left[\left\{
\omega/2
-\left(\eta(\boldsymbol{k})^2+|\mathcal{W}_0^{(\pm)}|^2
\right)^{1/2}
\right\}^2
\right.\nonumber\\
&&
\left.
+|\mathcal{V}_0^{(\pm)}|^2
\right]^{1/2}+n\omega,
\label{bandpn}
\end{eqnarray}
where
\begin{equation}
\eta(\boldsymbol{k})=d_4(\boldsymbol{k})-{\omega\over 2},
\end{equation}
and $d_4(\boldsymbol{k})$ is given in Eq.~(\ref{dk}).
Further,
$\mathcal{W}_0^{(+)}=[\mathcal{W}_0^{(-)}]^*=-i\Omega_x/2$ for the linearly polarized light, while
$\mathcal{W}_0^{(+)}= 0$ and $\mathcal{W}_0^{(-)}= i\Omega_c$ for the circularly polarized light.
Defining $\mathcal{V}_0^{\pm)}$ as an approximation expression of $\mathcal{V}^{(\pm)}$
of Eq.~(\ref{mathcalV2})
in view of the above approximations, one has
\begin{equation}
|\mathcal{V}_0^{(\pm)}|=t_{sp}\sigma^{(\pm)}\sqrt{\sin^2{(k_xd_x)}+\sin^2{(k_yd_y)}},
\end{equation}
where the prefactor $\sigma^{(\pm)}$ depending on the polarization of light 
is given by Eq.~(\ref{sigma+-}).
In Eqs.~(\ref{bandsn}) and (\ref{bandpn}), $\mathcal{V}_0^{(\pm)}$ is attributed to
$\mathcal{D}_3(\boldsymbol{k},t)$ and $\mathcal{D}_5(\boldsymbol{k},t)$ in Eq.~(\ref{H}),
while $\mathcal{W}_0^{(\pm)}$ is attributed to
$\mathcal{H}^\prime(t)$.
Hereafter, it is understood that in the opposite case that $hh(n+1)$ is located 
below $e(h-1)$, the subscript of $e(n-1)$ is replaced by that of $hh(n+1)$ in the above
equations, that is,
$E^{(\pm)}(\boldsymbol{k})=E^{(\pm)}_{hh(n+1)}(\boldsymbol{k}) \le n\omega$
and $E^{(\pm)}(\boldsymbol{k})=E^{(\pm)}_{e(n-1)}(\boldsymbol{k}) \ge n\omega$.
\begin{figure}[tb]
\begin{center}
  \includegraphics[width=8.5cm]{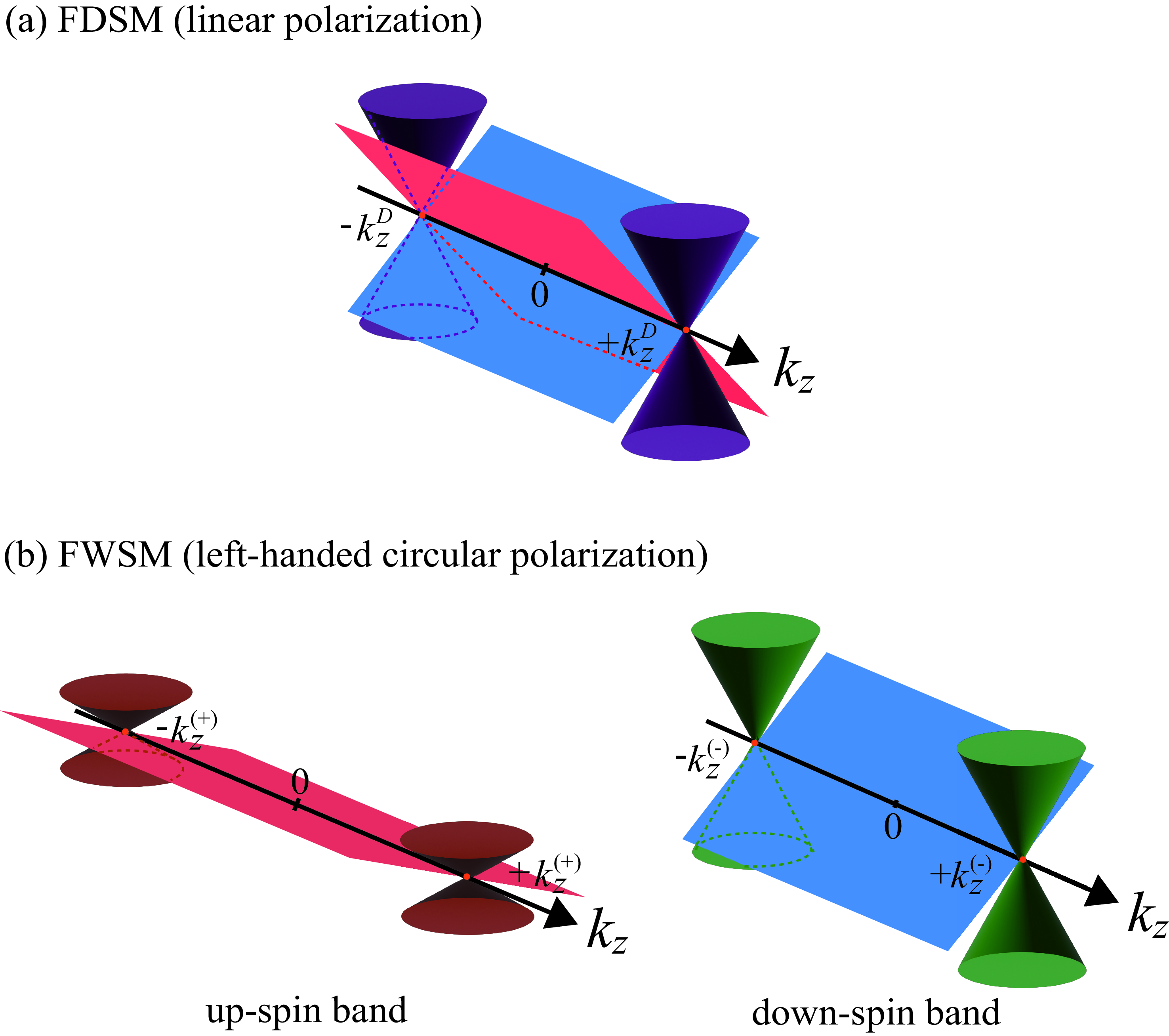}  
\caption{The scheme of the surface state formation. 
(a) In the FDSM arising from the drive of linearly polarized laser, 
this phase hosts
non-trivial surface states with up-spin (red) and down-spin (blue) that are pinned to
a pair of surface Dirac nodes at $k_z=\pm k_z^{D(\pm)}$.
(Here, $k_z^{D(\pm)}$ is replaced by the notation of $k_z^{D}$
just for the sake of simplicity.)
(b) In the FWSM arising from the drive of left-hand circularly polarized laser, 
due to the breaking of the T-symmetry, the Dirac nodes of the above FDSM are split into a pair of Weyl nodes,
and each FWSM phase hosts a non-trivial surface state.
One surface state (red) is characteristic of an up-spin band and 
is pinned to a pair of surface Dirac nodes at $k_z=\pm k_z^{(+)}$.
The other state (blue) is characteristic of a down-spin band and 
is pinned to a pair of surface Dirac nodes at $k_z=\pm k_z^{(-)}$.
(Here, $k_z^{W(\pm)}$ is replaced by the notation of $k_z^{(\pm)}$
just for the sake of simplicity.)
For more detail, consult the text.}
\label{fig1}
\end{center}
\end{figure}

Now, one examines the possibility of creating Dirac nodal points on the $k_z$-axis that result from band inversion for the irradiation of the linearly polarized light.
It is likely that the band $e(n-1)$ crosses the band $hh(n^\prime+1)$ for $n =n^\prime$
at $\boldsymbol{k}=(0,0,k_z)$
when 
$E_{e(n-1)}(\boldsymbol{k})=E_{hh(n+1)}(\boldsymbol{k})$.
Here, one takes account of the pair of Floquet bands of 
$e(-1)$ and $hh(1)$.
These bands are enabled to be inverted to form
a pair of Dirac nodes at the positions $\pm\boldsymbol{k}^{D(\pm)}\equiv \pm (0,0,k_z^{D(\pm)})$ subject to the equation
\begin{equation}
\epsilon_z(k_z^{D(\pm)})={1\over 2}\left[
\omega-E_{g}+
\sqrt{\omega^2-\Omega_x^2}
\right]
\label{Dirac}
\end{equation}
in terms of $\epsilon_z(k_z)$ defined in Eq.~(\ref{epsilonz0})
under the condition that
\begin{equation}
\omega_1^D<\omega<\omega_2^D,
\label{cond}
\end{equation}
where
\begin{equation}
\omega_1^D
=\text{Max}\left( \Omega_x,
\frac{\Delta^{2}_g+(\Omega_x/2)^2}
{\Delta_g}
\right),
\end{equation}
and
\begin{equation}
\omega_2^D
=
\frac{(\Delta_g-4t^z)^2+(\Omega_x/2)^2}
{\Delta_g-4t^z}.
\end{equation}
For more accurate expressions than Eqs.~(\ref{Dirac}) and (\ref{cond}), consult Eqs.~(\ref{node}) and (\ref{nodecond}), respectively.

The existence of these nodes exhibits the manifestation of the Floquet DSM (FDSM) phases
in the original crystal of Zn$_3$As$_2$ that is in a topologically trivial phase.
Due to the T-symmetry in addition with the time-glide I-symmetry, 
the up-spin and down-spin bands for the states $e(-1)$
and $hh(1)$ are doubly degenerate, that is,
$E^{(+)}_{e(-1)}(\boldsymbol{k})=E^{(-)}_{e(-1)}(\boldsymbol{k})$ and
$E^{(+)}_{hh(1)}(\boldsymbol{k})=E^{(-)}_{hh(1)}(\boldsymbol{k})$, and the above Dirac nodes are four-fold degenerate, namely, $k_z^{D(+)}=k_z^{D(-)}$.
Thus, it is considered that the FDSM carries Chern number zero and is not topologically protected.\cite{Armitage2018}
In addition, as shown in Eq.~(\ref{Dcone}),
$E^{(\pm)}(\boldsymbol{k})\ge0$ forms
an upper part of the Dirac cone (linear dispersion) in the vicinity of $\boldsymbol{k}=
\boldsymbol{k}^{D(\pm)}$, that is,
\begin{eqnarray}
&&E^{(\pm)}(\boldsymbol{k})\nonumber\\
&&\approx
\left[
\sum_{l=x,y,}(\xi^D_l)^2(k_ld_l)^2+(\xi^D_z)^2(\Delta k_zd_z)^2
\right]^{1/2},
\label{Dcone0}
\end{eqnarray}
where 
$\Delta k_z=k_z- k_z^{(D(\pm)}$, and the constants of 
$\xi^D_x=\xi^D_y$ and $\xi^D_z$ are given right below 
Eq.~(\ref{Dcone}).

Further, in the similar manner to DSMs created by the band inversion mechanism
in stationary systems such as Cd$_3$As$_2$ and Na$_3$Bi,\cite{Wang2012,Wang2013,Yang2014,Kargarian2016,Armitage2018}
two-dimensional (2D) nontrivial surface states are also expected in the FDSM phase.
As shown schematically in Fig.~\ref{fig1}(a),
these surface states are composed of up-spin and down-spin states forming a Kramers pair,
each energy band of which is attached to the same pair of the Dirac nodes that are projected 
to the surface 2D-BZ; hereafter, these projected Dirac nodes are termed as surface Dirac nodes or surface Dirac points.
The intersection of the Fermi energy with these two leaves of surface bands 
would result in the formation of double Fermi arcs, 
supposing that the whole of carriers are occupied just below $E_F$ 
in disregard of the non-equilibrium system
of concern.

Next, as regards the circularly polarized laser drive,
the T-symmetry is broken to lift the two-fold degeneracy between up-spin and down-spin bands.
Thus, the four-fold degeneracy at the Dirac nodes (at $\pm\boldsymbol{k}^{D(\pm)}$) are also lifted to be split into two pairs of Weyl nodes 
residing at $\pm\boldsymbol{k}^{W(-)}\equiv \pm (0,0,k_z^{W(-)})$ and
$\pm\boldsymbol{k}^{W(+)}\equiv \pm (0,0,k_z^{W(+)})$;
a double sign corresponds.
The nodal momentum $\pm\boldsymbol{k}^{W(-)}$ is attributed to the down-spin Floquet band,
and its location is subject to the similar equation as
Eq.~(\ref{Dirac}),
aside from the replacement of $\Omega_x$ by $2\Omega_c$,
\begin{equation}
\epsilon_z(k_z^{W(-)})={1\over 2}\left[
\omega-E_{g}+
\sqrt{\omega^2-4\Omega_c^2}
\right]
\label{Weyl-}
\end{equation}
under the condition that
\begin{equation}
\omega_1^{W(-)}<\omega<\omega_2^{W(-)},
\label{cond-}
\end{equation}
where
\begin{equation}
\omega_1^{W(-)}
=\text{Max}\left( 2\Omega_c,
\frac{\Delta^{2}_g+\Omega_c^2}
{\Delta_g}
\right),
\end{equation}
and
\begin{equation}
\omega_2^{W(-)}
=
\frac{(\Delta_g-4t^z)^2+\Omega_c^2}
{\Delta_g-4t^z}.
\end{equation}
Here the ac-Stark effect plays a key role.
On the other hand, 
the nodal momentum $\pm\boldsymbol{k}^{W(+)}$ is attributed to the up-spin Floquet band,
and its location is subject to the equation
\begin{equation}
\epsilon_z(k_z^{W(+)})=\omega-\Delta_{g}
\label{Weyl+}
\end{equation}
under the condition that
\begin{equation}
\omega_1^{W(+)}<\omega<\omega_2^{W(+)},
\label{cond+}
\end{equation}
where
$
\omega_1^{W(+)}=\Delta_{g}
$
and
$
\omega_2^{W(+)}=\Delta_{g}-4t^z.
$
Here, in contrast, the ac-Stark effect is less significant because 
the optical interaction given by $V^{(+)}(t)$ of Eq.~(\ref{Vcir})
has negligibly small contributions in the case of $\omega\approx E_g$;
consult Sec.~\ref{sec2B}.
Thus, it is obvious that $k_z^{W(+)}>k_z^{D(\pm)}>k_z^{W(-)}$;
based on Eqs.~(\ref{Dirac}), (\ref{Weyl-}) and (\ref{Weyl+}), approximate values of $k_z^{W(+)}$, $k_z^{D(\pm)}$, and $k_z^{W(-)}$
are estimated as 
$k_z^{W(+)}=0.956/d_z, k_z^{D(\pm)}=0.910/d_z$, and $k_z^{W(-)}=0.732/d_z$, respectively.

In addition, as shown in Eq.~(\ref{W-cone}),
$E^{(-)}(\boldsymbol{k})$ forms an upper part of
the Weyl cone (linear dispersion) in the vicinity of $\boldsymbol{k}=
\boldsymbol{k}^{W(-)}$, that is,
\begin{eqnarray}
&&E^{(-)}(\boldsymbol{k})\nonumber\\
&&\approx
\left[
\sum_{l=x,y,}(\xi^W_l)^2(k_ld_l)^2+(\xi^W_z)^2(\Delta k_zd_z)^2
\right]^{1/2},
\label{W-cone0}
\end{eqnarray}
where 
$\Delta k_z=k_z- k_z^{(W(-)}$, and the constants of 
$\xi^W_x=\xi^W_y$ and $\xi^W_z$ are given right below 
Eq.~(\ref{W-cone}).
On the other hand, 
as shown in Eq.~(\ref{W+cone}), in contrast wth $E^{(-)}(\boldsymbol{k})$,
$E^{(+)}(\boldsymbol{k})$ forms a quadratic dispersion with respect to $k_x$ and $k_y$ and a linear dispersion with respect to $\Delta k_z$ in the vicinity of $\boldsymbol{k}=
\boldsymbol{k}^{W(+)}$, that is,
\begin{eqnarray}
E^{(+)}(\boldsymbol{k})
\approx
\left|
(-t^{xy})\sum_{l=x,y}(k_ld_l)^2
+\eta_z^{(1)}(\Delta k_zd_z)
\right|,
\label{W+cone0}
\end{eqnarray}
where $\Delta k_z=k_z-k_z^{W(+)}$, and
$\eta^{(1)}_z=-2t^z\sin{(k_z^{W(+)}d_z)}$.
Here, a term of linear dispersion represented by 
$\sum_{l=x,y}\nu_l^{(+)}(k_ld_l)$ is considered negligibly small, because the expansion coefficient is given by $|\nu_l^{(+)}|=t_{sp}[t_{sp}J_1(z_c)/(2\eta(\boldsymbol{k}^{W(+)}))]^2$
due to the reduction of the spin-orbit interaction by the Peierls interaction by a factor of
the order of $J_1(z_c)$;
see also the explanation below Eq.~(\ref{W+cone}).

Further, the FWSM
band structure $E^{(+)}(\boldsymbol{k})$ in the vicinity of the $k_x-k_y$ plane
$(k_z=0)$ is examined.
Following Eq.~(\ref{bandpn}) for $d_4(\boldsymbol{k})>\omega/2$,
this is represented simply as 
\begin{equation}
E^{(+)}(\boldsymbol{k})
\approx \omega-d_4(\boldsymbol{k})
\end{equation}
within the order of $t_{sp}J_1(z_c)\approx 0$
due again to the reduction of $t_{sp}$ by the Peierls interaction.
Thus, there is a closed ring in the $k_x-k_y$ plane $(k_z=0)$ on which 
$E^{(+)}(\boldsymbol{k})\approx 0$; the locus of this ring is given by
\begin{eqnarray}
&&-2t^{xy}\sum_{l=x,y}[1-\cos{(k_ld_l)}]=\omega-\Delta_g >0.
\label{ring}
\end{eqnarray}
As regards the FWSM of the down-spin state, an effect of $\Omega_c$ causes 
a gap to open between 
$E^{(-)}_{hh(1)}(\boldsymbol{k})$ and
$E^{(-)}_{e(-1)}(\boldsymbol{k})$
in this plane.
Consult Eq.~(\ref{closedsurf}) and the related discussion in Appendix~\ref{app2-3}
on the closed ring formation in the $k_x-k_y$ plane $(k_z\not=0)$ and the 
origin of the difference between the up- and down-spin states.

Given the relation between Dirac points and Weyl points in stationary systems,
the surface Dirac point in the FDSM is regarded as
the stable merger of two Weyl points in the FWSM
that have different handedness and are projected to the same surface momentum.
Hereafter, these Weyl points are termed as surface Weyl nodes or surface Weyl points.
Due to the breaking of the T-symmetry and the resulting splitting of the Dirac node
into of the pair of Weyl nodes,
the associated energy bands of the two surfaces with different spin states are pinned to
different surface Weyl points,
as shown schematically in Fig.~\ref{fig1}(b).
That is,
the surface band characteristic of up-spin/down-spin state is pinned to the surface Weyl point 
projected from the bulk Weyl points at $\pm\boldsymbol{k}^{W(+)}/\pm\boldsymbol{k}^{W(-)}$.
Further, it is considered that the energy gap $\mathcal{E}_g^{(+)}$ arising from the hybridization between the up-spin Floquet bands $e(-1)$ and $hh(1)$ are largely
different from the energy gap $\mathcal{E}_g^{(-)}$ attributed to the down-spin bands
mostly due to the difference of magnitude between $V^{(+)}(t)$
and $V^{(-)}(t)$; to be more specific, $\mathcal{E}_g^{(+)} \ll \mathcal{E}_g^{(-)}$.
Such difference is straightforward reflected on the band gaps projected to the surface BZ;
see Fig.~\ref{fig1}(b).
Therefore, it is speculated that the most parts of down-spin surface band are energetically 
separated from the up-spin surface band; for more detail, consult Sec.~\ref{sec3C}.

\subsection{Floquet Band Structures of FDSM and FWSM}
\label{sec3B}

\begin{figure}[tb]
\begin{center}
\includegraphics[width=8cm,clip]{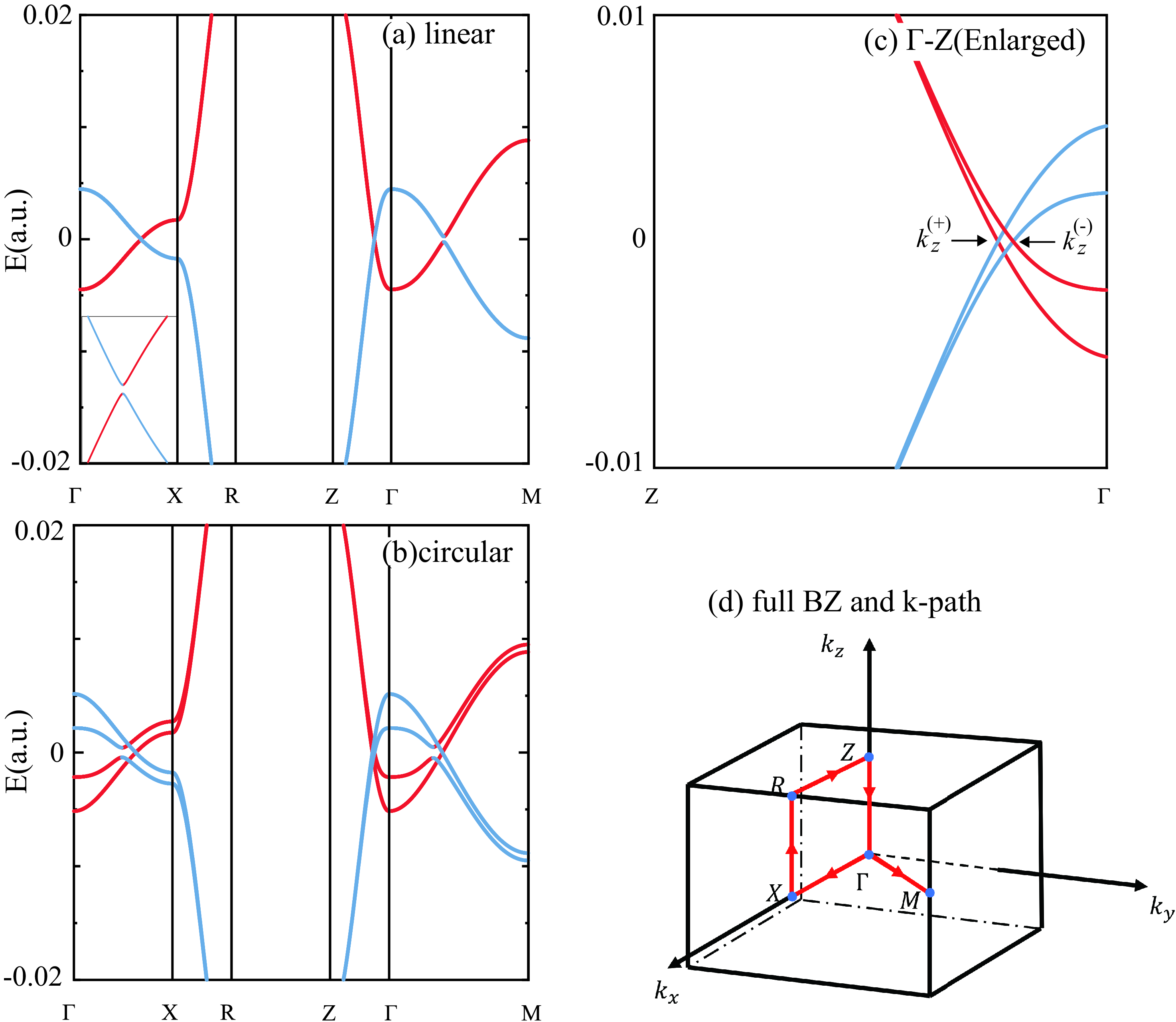}
\caption{Band structures of FDSM and FWSM.
(a) The calculated band structure of FDSM with the drive of a
linearly polarized laser. 
Inset: Expanded view of the band structure in the vicinity of the anticrossing along the
$\Gamma$-X line.
(b) The calculated band structure of FWSM with the drive of a left-hand
circularly polarized laser.
(c) Expanded view of panel (b) in the vicinity of the band crossing along the $\Gamma$-Z line
with specification of the Weyl nodes at $k_z^{(+)}$ and $k_z^{(-)}$;
these are the abbreviation of $k_z^{W(+)}$ and $k_z^{W(-)}$, respectively.
In panels (a), (b), and (c), bands dominated rather by the $s/p$-orbital component are denoted 
by a red/blue solid line.
(d) The bulk BZ of the crystal Zn$_3$As$_2$.
}
\label{fig2}
\end{center}
\end{figure}

Figures~\ref{fig2}(a) and \ref{fig2}(b) show the calculated Floquet band structures of FDSM and FWSM,
respectively, for the crystal structure of Zn$_3$As$_2$ given in Fig.~\ref{fig2}(d).
In Fig.~\ref{fig2}(a), it is found that there is a Dirac node along the $\Gamma$-Z line
at $k_z^{D(\pm)}$
in addition with anticrossings 
along the lines of $\Gamma$-X and $\Gamma$-M
with energy differences of approximately 3 and 30 meV, respectively.
As shown in Fig.~\ref{fig2}(b), the two-fold degeneracy confirmed in panel (a) is lifted to
result in energy splitting between the up-spin and down-spin bands.
It is noted that a pair of Weyl nodes emerges along the $\Gamma$-Z line
at different $k_z$'s following $k_z^{W(-)} < k_z^{W(+)}$, as shown in the enlarged view of
Fig.~\ref{fig2}(c).
As regards the up-spin bands, the anticrossing along the line of $\Gamma$-M 
is largely reduced
from that in panel (a) to approximately 0.4 meV, while the energy difference along the line of
$\Gamma$-X is almost the same as 2 meV.
In contrast, as regards the down-spin bands, the energy differences of anticrossings 
along the lines of $\Gamma$-X is largely enhanced from that in panel (a) to
approximately 23 meV, while that along the line of
$\Gamma$-M is slightly changed to approximately 20 meV.

\begin{figure}[tb]
\begin{center}
\includegraphics[width=8.5cm,clip]{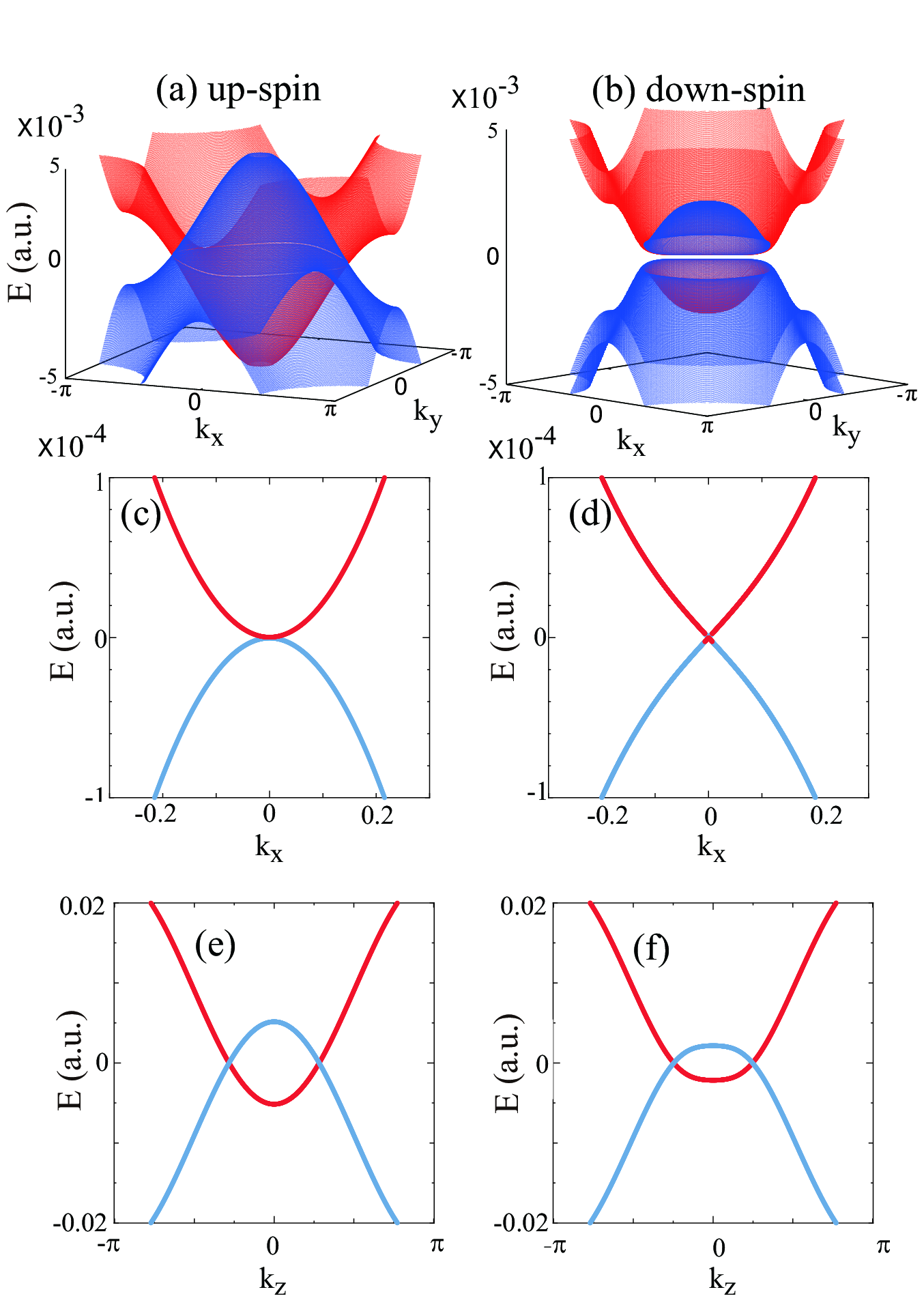}
\caption{Energy dispersions $E^{(\pm)}(\boldsymbol{k})$ of up-spin and down-spin bands at $k_z=0$
and in the vicinity of the Weyl points.
The axis of abscissa $k_l$ is gauged in the unit of $1/d_l$ with $l=x,y,z$.
Here, $E^{(\pm)}(\boldsymbol{k})$ is simply represented as
$E(\boldsymbol{k})$, and bands dominated rather by the $s/p$-orbital component are denoted 
by a red/blue solid line.
(a) $E(\boldsymbol{k})$ in the $k_x-k_y$ plane at $k_z=0$ for the up-spin bands.
(b) The same as panel (a) but for the down-spin bands.
(c) $E(\boldsymbol{k})$ as a function of $k_x$ with $k_y=0$ in the vicinity of the Weyl point $\boldsymbol{k}^{W(+)}$ for the up-spin bands.
(d) The same as panel (c) but in the vicinity of the Weyl point $\boldsymbol{k}^{W(-)}$
for the down-spin bands. 
(e) $E(\boldsymbol{k})$ as a function of $k_z$ with $k_x=k_y=0$ 
for the up-spin bands with the Weyl points $\pm k_z^{W(+)}=\pm 0.887/d_z$.
(f) The same as panel (e) but for the down-spin bands with
the Weyl points $\pm k_z^{W(-)}=\pm 0.775/d_z$. 
}
\label{fig3}
\end{center}
\end{figure}

Figures~\ref{fig3}(a) and \ref{fig3}(b) 
show the energy dispersions of the up-spin and down-spin bands at
$k_z=0$.
In Fig.~\ref{fig3}(a), the up-spin band structure is reminiscent of a NLSM phase 
with a nodal ring on the $k_x-k_y$ plane; see also the enlarged figure of it given in Appendix~\ref{app3}.
According to the analytic model developed in Sec.~\ref{sec3A},
the locus of the ring is approximately represented as Eq.~(\ref{ring}).
In fact, this ring is slightly blurred at most by $\mathcal{E}_g^{(+)}\approx 2$ meV
around $E=0$
that corresponds to the above-mentioned energy difference
along the line of $\Gamma$-X.
On the other hand, it is obviously seen in Fig.~\ref{fig3}(b)
that the down-spin band is gapped by the order of $\mathcal{E}_g^{(-)}\approx 20$ meV
due to the relatively strong anticrossing between $e(-1)$ and $hh(1)$; see also the enlarged figure of it given in Appendix~\ref{app3}.

The definite contrast in the energy dispersions between up-spin and down-spin bands
seen in Figs.~\ref{fig3}(a) and \ref{fig3}(b)
is caused by the different manner of couplings between the Floquet states of 
$e(-1)$ and $hh(1)$.
It is noted that the leading contribution arises from a two-photon coupling
between $e(-1)$ and $hh(1)$, 
because the difference of
the photon number of these Floquet bands equals two.
This coupling is given by a successive interaction composed of the coupling 
due to $\tilde{D}_{4,nn^\prime}(\boldsymbol{k},\omega)$ and 
one of the three terms, $\tilde{D}_{j,nn^\prime}(\boldsymbol{k},\omega), (j=3, 5)$
and $\tilde{\mathcal{H}}^\prime_{nn^\prime}(\omega)$, in Eq.~(\ref{Ltilde})
with $|n-n^\prime|=1$.
For example, for down-spin bands,
it is likely that $hh(1)$ is mediated by a two-photon interaction,
$\tilde{\mathcal{H}}^\prime_{01}(\omega)$ followed by $\tilde{D}_{4,-10}(\boldsymbol{k},\omega)$, to be coupled with $e(-1)$.
As regards up-spin bands, because the effect of $\tilde{\mathcal{H}}^\prime_{01}(\omega)$
is negligibly small,
$hh(1)$ is couplied with $e(-1)$ by a two-photon interaction,
$\tilde{D}_{j,01}(\boldsymbol{k},\omega), (j=3, 5)$ followed by $\tilde{D}_{4,-10}(\boldsymbol{k},\omega)$.
The magnitudes of interactions $\tilde{\mathcal{H}}^\prime_{01}(\omega)$
and $\tilde{D}_{j,01}(\boldsymbol{k},\omega), (j=3, 5)$ are roughly evaluated to be
$\Omega_c$ and $J_1(z_c)t_{sp}$, respectively;
consult Eqs.~(\ref{Vcir}), (\ref{appD3}), and (\ref{appD5}).
Thus, it is stated that the inverted band gap $\mathcal{E}_g^{(-)}$ in the down-spin state is mostly caused by
a strong resonant electric dipole 
coupling, while $\mathcal{E}_g^{(+)}$ in the up-spin state is just
attributed to an optical spin-orbit coupling, namely, a spin-orbit coupling reduced by a factor of $J_1(z_c)$; actually, $\Omega_c=5.49\times10^{-3} \gg t_{sp}J_1(z_c)=2.07\times10^{-4}$.
The resulting FWSM phase for the up-spin state is considered as a Floquet NLSM (FNLSM)
phase that is generated by the drive of the present circularly polarized light.

Such sharp distinction between the up-spin and down-spin bands is also seen
in the energy dispersions in the $k_x$-direction
in the vicinity of the Weyl points at $\boldsymbol{k}^{W(+)}$ and $\boldsymbol{k}^{W(-)}$,
respectively, as shown in Figs.~\ref{fig3}(c) and \ref{fig3}(d).
It is noted that the up-spin band crosses with quadratic band touching, 
following 
\begin{equation}
E^{(+)}(\boldsymbol{k})
\approx
(-t^{xy})(k_xd_x)^2\ge 0,
\label{W+cone1}
\end{equation}
as given in Eq.~(\ref{W+cone0}),
while the down-spin band crosses with linear band touching, as often happens, following
\begin{equation}
E^{(-)}(\boldsymbol{k})
\approx
\xi^W_x|k_xd_x|\ge 0,
\label{W-cone1}
\end{equation}
as given in Eq.~(\ref{W-cone0}),
where $\xi^W_x$ is of the order of $t_{sp}$.
In fact, there is a contribution from the linear dispersion of the form of $\nu_x^{(+)}(k_xd_x)$
in Eq.~(\ref{W+cone1}), however, this is neglected because 
of $\nu_x^{(+)}\ll 1$, as mentioned before.

Further, it is seen in Figs.~\ref{fig3}(e) and \ref{fig3}(f) that
the two bands of $e(-1)$ and $hh(1)$ are inverted to form a pair of Weyl nodes along
the $k_z$-axis at $k_z^{W(\pm)}$ and $-k_z^{W(\pm)}$.
These dispersions in the vicinity of $\boldsymbol{k}^{W(\pm)}$ are given by
\begin{equation}
E^{(+)}(\boldsymbol{k})
\approx
\eta_z^{(1)}|\Delta k_zd_z|\ge 0
\label{W+cone3}
\end{equation}
for the up-spin state,
and
\begin{equation}
E^{(-)}(\boldsymbol{k})
\approx
\xi^W_z|\Delta k_zd_z|\ge 0
\label{W-cone3}
\end{equation}
for the down-spin state, following
Eqs.~(\ref{W+cone0}) and (\ref{W-cone0}), respectively.
Here, the obtained numerical value of
$k_z^{W(+)}(=0.887/d_z)$ is greater than that of $k_z^{W(-)}(=0.775/d_z)$, which is in harmony with
the qualitative discussion based on the approximated expressions of
Eqs.~(\ref{Weyl-}) and (\ref{Weyl+}); consult the values thus obtained for
$k_z^{W(\pm)}$ below Eq.~(\ref{cond+}).
It is speculated that the difference of the former numerical values from the latter approximate ones is attributed to the non-resonant contributions of interband couplings
beyond the rotational-wave approximation in
the nearly resonant two-band model adopted in Sec.~\ref{sec3A}.
Actually, the intense laser field is applied to the system of concern with the order of $\Omega_c/\omega\approx 0.37$, and hence, for instance, the Floquet band $hh(1)$ is 
somewhat
coupled with other non-resonant bands of $e(n\not=0)$
in addition with the nearly resonant band $e(0)$.

\subsection{Surface States}
\label{sec3C}

Here, it is considered that a vanishing boundary condition in the $y$-direction is imposed on the Floquet eigenvalue problem given by Eq.~(\ref{Floquet}) in place of a periodic boundary condition.
To be specific, an electron is confined in the finite range of $y$ from $L_1=0$ to $L_2=40$ a.u., while it moves freely in the $x-z$ plane.
Such confinement results in energy dispersions $\mathcal{E}(\bar{\boldsymbol{k}})$ 
that are the projection of bulk bands $E(\boldsymbol{k})$ on the $k_x-k_z$ plane
where $\bar{\boldsymbol{k}}=(k_x,k_z)$.
Further, it is likely that surface states are hosted by the projected bands.
For the sake of the later convenience, the positions of surface Weyl nodes
for the up-spin and down-spin bands are
represented as $\pm\bar{\boldsymbol{k}}^{W(+)}=\pm(0,k_z^{W(+)})$ and
$\pm\bar{\boldsymbol{k}}^{W(-)}=\pm(0,k_z^{W(-)})$, respectively.

\begin{figure}[t]
\begin{center}
\includegraphics[width=8cm,height=9cm]{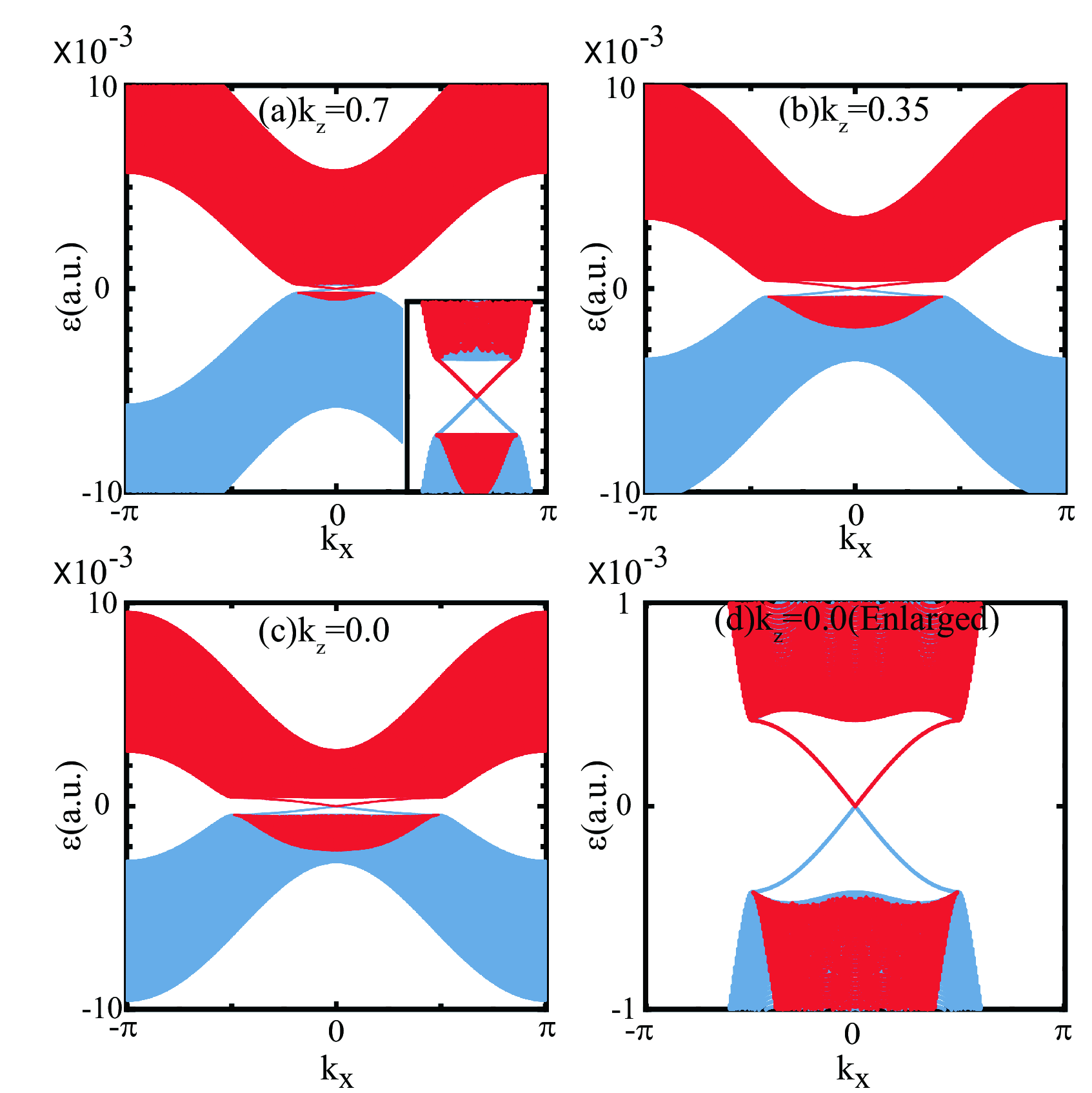}
\caption{Projected energy dispersions $\mathcal{E}(\bar{\boldsymbol{k}})$ with a surface state of down-spin.
The axis of abscissa $k_x$ is gauged in the unit of $1/d_x$.
Here, bands dominated rather by the $s/p$-orbital component are denoted 
by a red/blue solid line.
(a) $\mathcal{E}(\bar{\boldsymbol{k}})$ at $k_z=0.7$ slightly smaller than $k_z^{W(-)}$.
(b) The same as pannel (a) but at $k_z=0.35$.
(c) The same as pannel (a) but at $k_z=0$.
(d) Enlarged view of pannel (c).
}
\label{fig4}
\end{center}
\end{figure}

Figure~\ref{fig4} shows the projected energy dispersions of down-spin bands with surface states at three different $k_z$'s.
As shown in Fig.~\ref{fig4}(a), at $k_z$ close to $k_z^{W(-)}$,
the inverted bands of $e(-1)$ and $hh(1)$ form a definite energy gap $\mathcal{E}_g^{(-)}$,
hosting a pair of surface states just in a small range of $k_x$.
It is evident that as $k_z$ becomes closer to $k_z^{(-)}$, the range of $k_x$ becomes more reduced, and eventually, the pair of surface states are embedded in the surface Weyl point
at $\bar{\boldsymbol{k}}^{W(-)}$.
Meanwhile, it is remarked that the appearance of such a pair is due to a numerical artifact ascribable to the above-mentioned confinement
of electron in the finite range in place of a semi-infinite confinement corresponding to
$L_2=\infty$.
Here, it is understood that in all of the figures in Fig.~\ref{fig4}, just the surface states with a positive gradient are taken account of.
As shown in Figs.~\ref{fig4}(b)-\ref{fig4}(d), with the further decrease of $k_z$,
the range of $k_x$ in which the surface state is supported becomes larger, and is maximized 
at $k_z=0$, where this range extends over a half of the BZ in the $k_x$ direction.
Moreover, as $k_z$ changes from $k_z=0$ to the negative $k_z$-direction, the range of $k_x$ turns to a decrease, and eventually, 
at $k_z=-k_z^{W(-)}$, the surface state is incorporated with another surface Weyl point at
$-\bar{\boldsymbol{k}}^{W(-)}$; though not shown here.
These nontrivial surface states sliced in the interval $-k_z^{W(-)}\le k_z \le k_z^{W(-)}$ are unified to form a tilted surface band in the $k_x-k_z$ plane.
Both edges of it are pinned to the respective surface Weyl points at $\pm \bar{\boldsymbol{k}}^{W(-)}$.
This surface band is schematically depicted
as the tilted surface that is colored blue in the right figure of Fig.~\ref{fig1}(b).

\begin{figure}[tb]
\begin{center}
\includegraphics[width=8cm,height=9cm]{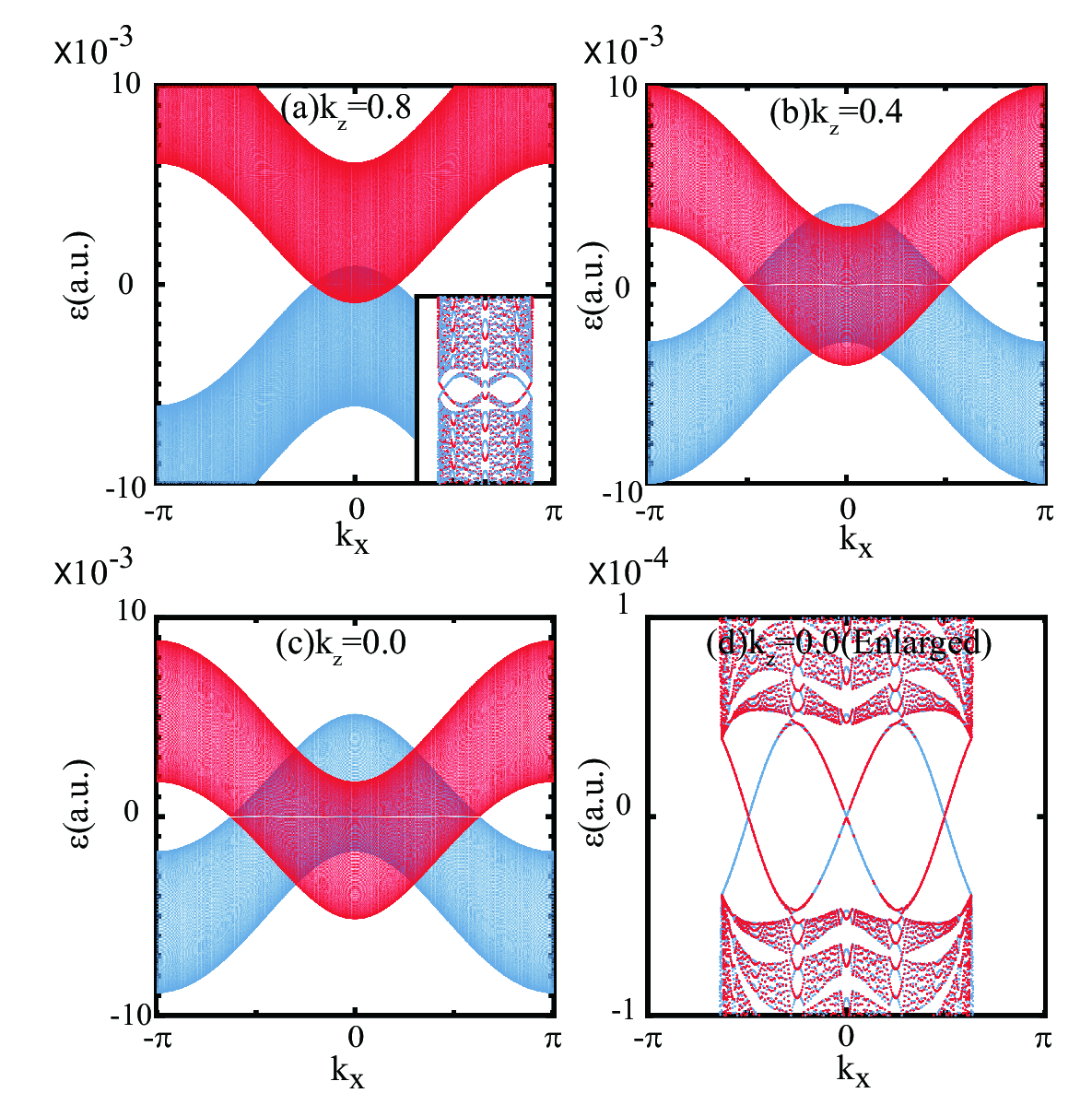}
\caption{Projected energy dispersions $\mathcal{E}(\bar{\boldsymbol{k}})$ with a surface state of up-spin.
The axis of abscissa $k_x$ is gauged in the unit of $1/d_x$.
Here, bands dominated rather by the $s/p$-orbital component are denoted 
by a red/blue solid line.
(a) $\mathcal{E}(\bar{\boldsymbol{k}})$ at $k_z=0.8$ slightly smaller than $k_z^{W(+)}$.
(b) The same as pannel (a) but at $k_z=0.4$.
(c) The same as pannel (a) but at $k_z=0$.
(d) Enlarged view of pannel (c).
}
\label{fig5}
\end{center}
\end{figure}

Figures~\ref{fig5}(a)-\ref{fig5}(c) show the projected energy dispersions of up-spin bands with surface states at three different $k_z$'s.
It is seen that the pattern of variance of the surface states formed here follows that shown in the down-spin bands of Fig.~\ref{fig4}.
However, the energy gap $\mathcal{E}_g^{(+)}$
is extremely smaller than $\mathcal{E}_g^{(-)}$, and as seen in Fig.~\ref{fig5}(d), the surface band is 
slightly tilted with overall negative gradient and undulation.
The pattern of variance in the negative $k_z$-direction is also
subject to that seen in the down-spin bands; though not shown here.
As a result, the nontrivial surface states sliced in the interval $-k_z^{W(+)}\le k_z \le k_z^{W(+)}$ form a slightly tilted and undulated surface band in the $k_x-k_z$ plane.
Both edges of it are pinned to the respective surface Weyl points
at $\pm\bar{\boldsymbol{k}}^{W(+)}$.
This surface band is schematically depicted
as the tilted surface that is colored red in the left figure of Fig.~\ref{fig1}(b).

\subsection{Physical Properties}
\label{Sec3D}

First, discussion is made on the chirality of the FWSM phases and the related topological phase transitions.
It is considered that based on the qualitative discussion in Sec.~\ref{sec3A}, the conditions of generating the Weyl nodes for the up-spin and down-spin states are approximately evaluated as Eqs.~(\ref{cond+}) and ~(\ref{cond-}),
respectively.
According to these, when $\omega$ is made greater from $\omega_1^{W(\pm)}$ and eventually identical to $\omega_2^{W(\pm)}$,
the Weyl nodes at $\boldsymbol{k}^{W(\pm)}$ move along $k_z$ axis from the $\Gamma$ point $k_z=0$ toward the boundary of the BZ 
at $k_z=\pi/d_z$ to annihilate with the other pair of the Weyl nodes at $-\boldsymbol{k}^{W(\pm)}$ that move in the opposite direction toward the boundary
at $k_z=-\pi/d_z$; a double sign corresponds.
This implies that the Weyl nodes at $\boldsymbol{k}^{W(\pm)}$ possess
opposite handedness from that at the other Weyl nodes at $-\boldsymbol{k}^{W(\pm)}$.
In other words, there should be the relations that 
\begin{equation}
h_+^{(+)}h_-^{(+)} =-1,\;h_+^{(-)}h_-^{(-)} =-1, 
\label{hrel1}
\end{equation}
where
$h_{\pm}^{(+)}$ and $h_{\pm}^{(-)}$, which are either 1 or -1, represent helicities of the Weyl cones at 
$\pm\boldsymbol{k}^{W(+)}$ for the up-spin and $\pm\boldsymbol{k}^{W(-)}$ for the down-spin, respectively.
Further, it is noted that the handedness of the Weyl node for the up-spin state
at $\boldsymbol{k}^{W(+)} (-\boldsymbol{k}^{W(+)})$ is opposite from that for the down-spin
state at $\boldsymbol{k}^{W(-)} (-\boldsymbol{k}^{W(-)})$, because
a pair of Weyl nodes for the up-spin and down-spin states at $\boldsymbol{k}^{W(+)}$ and $\boldsymbol{k}^{W(-)}$, respectively, are generated by splitting of the Dirac node at $\boldsymbol{k}^D(-\boldsymbol{k}^D)$ due to the breaking of the T-symmetry. 
That is, there should be the relations that
\begin{equation}
h_+^{(+)}h_+^{(-)} =-1, \;h_-^{(+)}h_-^{(-)} =-1.
\label{hrel2}
\end{equation}
Actually, the above relations of Eqs.~(\ref{hrel1}) and (\ref{hrel2}) are confirmed 
by defining these helicities as Eqs.~(\ref{h++}) and (\ref{h+-}), followed by
mathematical evaluation, as developed in Appendix~\ref{app2-4}.
Here, these expressions of helicities are extracted from effective Fourier-Weyl Hamiltonians 
of Eqs.~(\ref{LW+}) and (\ref{LW-}), which are 
reduced from the original Floquet Hamiltonian of Eq.~(\ref{Ltilde}).
Here, one mentions that recently, dynamical characterization of Floquet-Weyl nodes is discussed in Ref. \onlinecite{Umer2021b}.

In passing, when $\omega$ exceeds $\omega_2^{W(\pm)}$, the topological order is changed from the FWSM phase to a phase of Floquet topological insulator
due to the gap opening.
Further, the reduction of $\omega$ below $\omega_1^{W(\pm)}$ in the other direction 
brings the FWSM phase just back to a trivial insulator phase.

\begin{figure}[tb]
\begin{center}
\includegraphics[width=8cm,height=9cm]{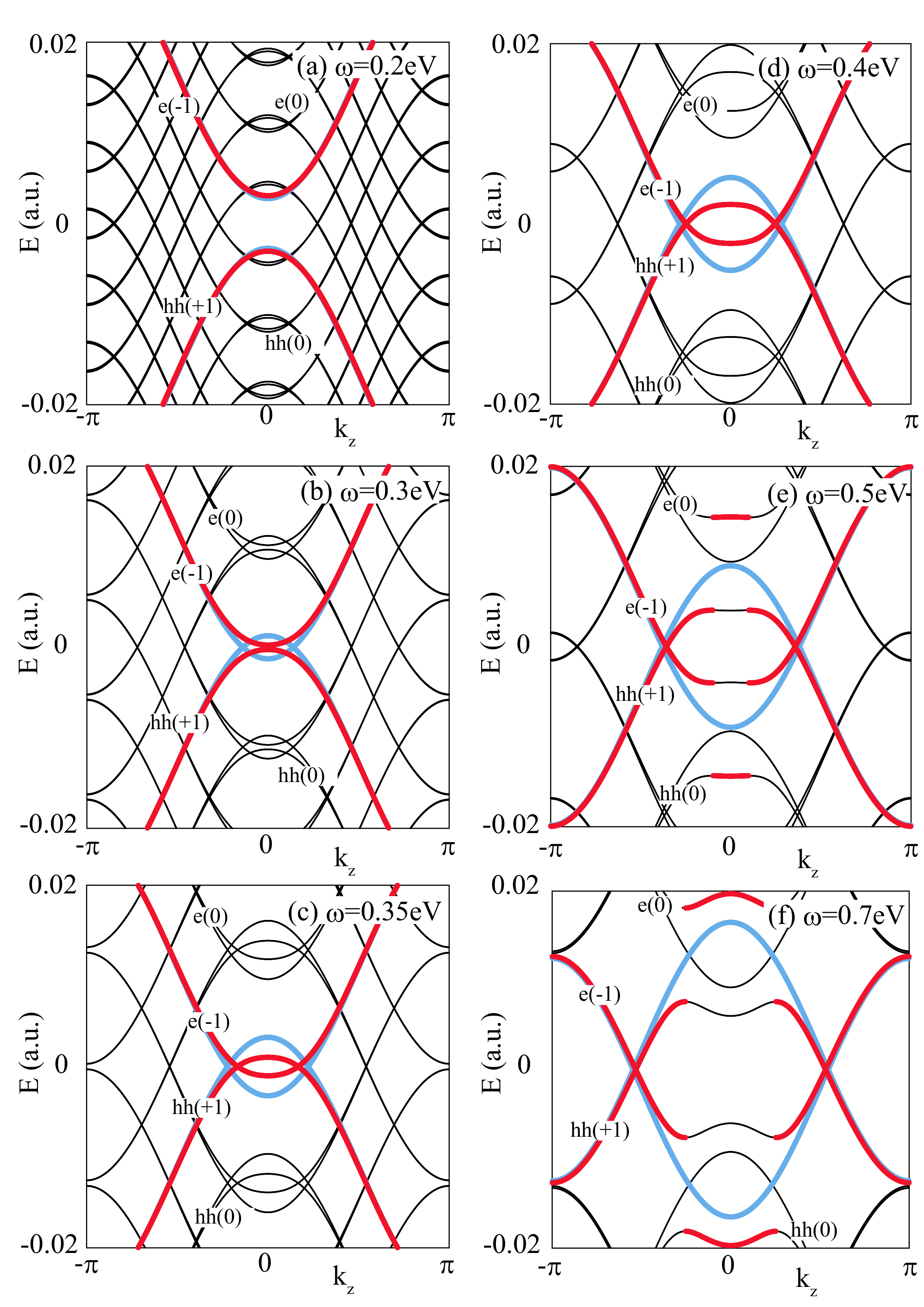}
\caption{Frequency dependence of energy dispersions $E^{(\pm)}(\boldsymbol{k})$ 
of up-spin and down-spin bands as a function of $k_z$ with $k_x=k_y=0$ for $\omega$
equal to
(a) 0.2\:eV (7.35$\times 10^{-3}$ a.u.),
(b) 0.3\:eV (1.10$\times 10^{-2}$ a.u.),
(c) 0.35\:eV (1.29$\times 10^{-2}$ a.u.),
(d) 0.4\:eV (1.47$\times 10^{-2}$ a.u.),
(e) 0.5\:eV (1.84$\times 10^{-2}$ a.u.), and
(f) 0.7\:eV (2.57$\times 10^{-2}$ a.u.).
The abscissa $k_z$ is gauged in the unit of $1/d_z$.
The panel (d) is the same as Figs.~\ref{fig3}(e) and \ref{fig3}(f).
Here, the legend of ordinate $E^{(\pm)}(\boldsymbol{k})$ is simply represented as
$E$.
The up-spin bands $e(-1)$ and $hh(1)$ are depicted 
by blue solid lines, and the down-spin bands $e(-1)$ and $hh(1)$ are 
depicted by red solid lines.
The parent bands are also labeled as $e(0)$ and $hh(0)$.
In all panels, red solid lines are partially superimposed on blue solid lines.
}
\label{fig6}
\end{center}
\end{figure}

Figure~\ref{fig6} shows the energy dispersions $E^{(\pm)}(\boldsymbol{k})$ 
of up-spin and down-spin bands as a function of $k_z$ with $k_x=k_y=0$ for $\omega=0.2\sim 0.7$ eV.
Below, discussion is made on the alteration of just $e(-1)$ and $hh(1)$ bands
of an up-spin state (depicted by blue solid lines) and a down-spin state
(depicted by red solid lines) with respect to $\omega$.
In the panel (a) for $\omega=0.2$ eV, both spin bands are not inverted, 
and in the panel (b) for $\omega=0.3$ eV, 
the up-spin band is inverted 
to form a FWSM phase with a pair of Floquet-Weyl nodes
while the down-spin band is left open.
In the panel (c) for $\omega=0.35$ eV, both bands become inverted
to form FWSM phases with two pairs of Floquet-Weyl nodes, 
and similarly, in the panel (d) for $\omega=0.4$ eV $(< E_g$), 
both bands remain inverted.
Such band inversion is still retained in the panels (e) and (f) even for $\omega=$0.5 and 0.7 eV $(> E_g$).
Incidentally, the discontinuities of the down-spin band
seen in the panels (e) and (f) are due to an anticrossing between 
$e(-1)$ and $hh(0)$ and that between 
$e(0)$ and $hh(1)$.
It is worth comparing these numerical results with the results estimated 
by Eqs.~(\ref{cond+}) and (\ref{cond-})
based on the analytic model in Sec.~\ref{sec3A},
where $\omega_1^{W(+)}=0.23\: {\text eV}, \omega_2^{W(+)}=0.64\: {\text eV}, \omega_1^{W(-)}=0.33 \:{\text eV}$, and $\omega_2^{W(-)}=0.67\: {\text eV}$.
It is found that aside from the panel (f), 
the above-stated changes of topological order with respect to $\omega$ are 
well consistent with these estimated existence conditions of FWSM phases.
The variance seen in the panel (f) is due to the breaking of the rotational-wave approximation adopted in this model.
Actually, this approximation is considered accurate under the situation
that $\omega \approx E_g (= 0.46$ eV).

Second, discussion is made on a magnetic property induced by the irradiation of the 
intense laser with a left-hand circular polarization.
As far as the nearly resonant optical transition is concerned,
down-spin electrons that are situated in a valence band before the irradiation are selectively excited to a conduction band, and some fractions of the excited 
electrons are deexcited back to the valence band due to
the Rabi oscillation, 
whereas up-spin electrons remain almost in the valence band;
consult Sec.~\ref{sec2B}.
In terms of the Floquet picture,
these excitation and deexcitation processes in a series of the non-equilibrium dynamics
are interpreted as couplings between 
one pair of down-spin bands $hh(1)$ and $e(0)$, and between
another pair of down-spin bands $hh(0)$ and $e(-1)$, respectively.
Thus, carriers are likely distributed to both bands of $e(-1)$ and $hh(1)$,
which are further coupled by the two-photon interaction mentioned in
Sec.\ref{sec3B} to form the FWSM phase through the ac-Stark splitting.
On the contrary, it is considered that the up-spin bands of $e(-1)$ and $hh(1)$ are almost unoccupied.
Therefore, the down-spin electrons are exclusively distributed over the surface, 
while these
coexist with the up-spin electrons in the bulk though both electronic states are 
energetically separated by the amount of $\mathcal{E}_g^{(-)}$.

This implies that the system of concern exhibits transient surface
magnetization with down spins that survives for
as long as the associated population relaxation time,
besides bulk magnetization that is expected to be induced as well.
In addition with such an effect of spin magnetization, it is likely that
the circularly polarized laser induces the inverse Faraday effect, which is
a sort of a generation mechanism of orbital magnetization.
\cite{Pershan1966,Kimel2005,Hertel2006,Zhang2009,Battiato2014}
This effect is expected to contribute the above surface magnetization to a certain extent.
Indeed, the surface magnetization seems faint and transient, but the magnitude of it can
be somewhat enhanced by increasing the strength of the circularly polarized laser.
Moreover, the measurement of such an intriguing phenomenon would be feasible
by means of the longitudinal magneto-optic Kerr effect that can detect the degree of strength of magnetization manifested just in the surface.
\cite{Kerr1877,Weinberg2008,Haider2017}
To do this, a pump-probe measurement is expected to be effective, in which
a linear polarized laser causing the magneto-optic Kerr effect is 
incorporated as a probe in addition with the pump laser with the left-hand circular 
polarization.

Below, additional comments
on the results described in Secs~\ref{sec3A}-\ref{sec3C} are enumerated.

(1) The up-spin FWSM band is considered as a FNLSM phase, because
the band gap of $\mathcal{E}_g^{(+)}$ is of the order of 2 meV; consult Sec~\ref{sec3B}.
Actually, such small energy separation and
the concomitant surface state would be possibly
smeared with homogeneous broadening due to an electron correlation effect
and inhomogeneous broadening due to finite temporal width of a laser pulse
--- in place of the ideal continuous-wave laser ---, which is of the order of
a couple of meV for a pico-second pulse.
Although the up-spin bands are almost unoccupied as stated above,
these would be detectable by reconstructing the optical system of concern as follows:
the up-spin bands of $e(-1)$ and $hh(1)$ are excited in advance by an intense ultrashort pulse laser with linear polarization, followed by the irradiation of the pico-second pulse (the continuous-wave laser) with the left-hand circular polarization.

(2) In view of the above comment (1), the surface states hosted by the down-spin band are entirely embedded in 
the continuum (bulk) of the FNLSM phase of the up-spin band; consult Figs.~\ref{fig4} and \ref{fig5}.
When a spin flip interaction attributed to the spin-orbit coupling is tuned on, the surface states become somewhat unstable 
due to the effect of Fano resonance, namely, the collapse of the discrete levels of the surface states into the continuum states which is caused by the interaction between both of these states.\cite{Fano1961}
The spin flip interaction becomes effective when either $d_1(\boldsymbol{k})$ or $d_2(\boldsymbol{k})$ has a non-negligible contribution to 
the effective Hamiltonian given in Eq.~(\ref{Hcal2}).

(3) The crystal Zn$_3$As$_2$ has a bulk rotational symmetry around the $z$-axis,
and this leads to the formation of FWSMs under the conditions of Eqs.~(\ref{cond-})
and (\ref{cond+}).
In fact, there remains internal compression normal to this axis within the crystal, 
and this symmetry is considered partially broken.
Such breaking will open up a slight gap to make the Floquet system of concern
insulating.\cite{Wang2012}

\section{Conclusions}
\label{sec4}

It is found that the narrow gap semiconductor Zn$_3$As$_2$ is driven by
a left-hand circularly-polarized continuous-wave laser with 
frequency nearly resonant with the band gap $E_g$
to produce the two types of FWSM phases simultaneously in the crystal,
which are sharply distinguished by their spins.
The bulk rotational symmetry around the $z$-axis protects a pair of Weyl nodes 
with opposite chirality along the
$k_z$-axis in the respective FWSM phases
under the condition of either Eq.~(\ref{cond-}) or Eq.~(\ref{cond+}).
In the down-spin FWSM phase, the Floquet bands of $e(-1)$ and $hh(1)$ touch
in a linear manner
in the vicinity of the Weyl nodes situated 
at $\pm\boldsymbol{k}^{W(-)}$, hosting the 
nontrivial surface states pinned to both nodes.
Since the above-mentioned laser makes electrons excited exclusively
in the down-spin Floquet bands, it is considered that
the surface states are selectively occupied by
such spin-polarized electrons, showing transient magnetization with
partial modification by the inverse Faraday effect.
This surface magnetization would be measured by virtue of the magneto-optic Kerr effect.
On the other hand, in the up-spin FWSM phase, 
the Floquet bands of $e(-1)$ and $hh(1)$ touch in the vicinity of the Weyl nodes situated 
at $\pm\boldsymbol{k}^{W(+)}$
almost in a quadratic manner in the $k_x$- and $k_y$-directions and in a linear manner
in the $k_z$-direction.
Because of the negligibly small band gap, this up-spin FWSM phase is rather
considered as the FNLSM phase.
To detect this phase somehow or other,
it would be necessary to make excited electrons occupied in the up-spin bands in advance
prior to the irradiation of the circularly polarized laser.
The exploration of the transient non-equilibrium dynamics of the concerned system
is inevitable in addition with Floquet band structures
to deepen the understandings of the underlying physics of the FWSMs.


\begin{acknowledgments}
This work was supported by JSPS KAKENHI Grant No. JP19K03695.
The authors are grateful to Prof. J. Fujioka for fruitful comments and discussion.
\end{acknowledgments}

\appendix
\section{Explicit expressions of 
$\tilde{D}_{j,nn^\prime}(\boldsymbol{k},\omega)\;\;(j=3\sim5)$
and
$\tilde{V}^{(\pm)}_{nn^\prime}$}
\label{app1}

The Floquet matrix element $\tilde{D}_{j,nn^\prime}(\boldsymbol{k},\omega)$
seen in Eq.~(\ref{Ltilde}) is given by
\begin{equation}
\tilde{D}_{j,nn^\prime}(\boldsymbol{k},\omega)
={1\over T}\int_{0}^T dt \:e^{-i\Delta n\omega t} 
D_j(\boldsymbol{k},t)
\end{equation}
with $\Delta n=n-n^\prime$.
This is expressed in terms of the $N$th-order Bessel function of the first kind
\begin{equation}
J_N(z_l)={1\over 2\pi}\int_{0}^{2\pi} d\theta \:e^{-iN\theta} e^{iz_l\sin{\theta}}
\label{JN}
\end{equation} 
with $z_l=F_ld_l/\omega\;\;(l=x,y)$
as follows:
\begin{eqnarray}
\tilde{D}_{3,nn^\prime}(\boldsymbol{k},\omega)
=
\left\{
\begin{array}{l}
t_{sp}J_{\Delta n}(z_x)\sin{(k_x d_x)}\\
\hspace{9mm}{\rm for}\;\;\Delta n=0,\pm2,\pm4,\cdots\\
it_{sp}J_{\Delta n}(z_x)\cos{(k_x d_x)}\\
\hspace{9mm}{\rm for}\;\;\Delta n=\pm1,\pm3,\cdots
\end{array}
\right.,
\label{appD3}
\end{eqnarray} 
\begin{eqnarray}
&&\tilde{D}_{4,nn^\prime}(\boldsymbol{k},\omega)\nonumber\\
&&=
\left\{
\begin{array}{l}
\Delta_{g}
-\Sigma_{l=x,y}2t^{xy}\left[1-J_{0}(z_l)\cos{(k_l d_l)}\right]\\
\hspace{9mm}-2t^{z}\left[1-\cos{(k_z d_z)}\right]\\
\hspace{3cm}{\rm for}\;\;\Delta n=0\\

2t^{xy}\left[J_{\Delta n}(z_x)\cos{\left(k_x d_x\right)}\right.\\
\hspace{9mm}
\left. +J_{\Delta n}(z_y)\cos{\left(k_y d_y+{\Delta n\over 2}\pi\right)}\right]\\
\hspace{3cm}{\rm for}\;\;\Delta n=\pm2,\pm4,\cdots\\

2t^{xy}\left[-iJ_{\Delta n}(z_x)\sin{\left(k_x d_x\right)} \right.\\
\hspace{9mm}
\left. +J_{\Delta n}(z_y)\cos{\left(k_y d_y+{\Delta n\over 2}\pi\right)}\right]\\
\hspace{3cm}{\rm for}\;\;\Delta n=\pm1,\pm3,\cdots\\

\end{array}
\right.,
\label{appD4}
\end{eqnarray} 
and
\begin{eqnarray}
\tilde{D}_{5,nn^\prime}(\boldsymbol{k},\omega)
=
\left\{
\begin{array}{l}
i^{\Delta n}t_{sp}J_{\Delta n}(z_y)\sin{(k_y d_y)}\\
\hspace{9mm}{\rm for}\;\;\Delta n=0,\pm2,\pm4,\cdots\\
i^{(\Delta n-1)}t_{sp}J_{\Delta n}(z_y)\cos{(k_y d_y)}\\
\hspace{9mm}{\rm for}\;\;\Delta n=\pm1,\pm3,\cdots
\end{array}
\right..
\label{appD5}
\end{eqnarray} 

Further, the Fourier transform of the electric-dipole interaction
$V^{(\pm)}(t)$ of Eq.~(\ref{V+-}), given by
\begin{equation}
\tilde{V}^{(\pm)}_{nn^\prime}
={1\over T}\int_{0}^T dt \:e^{-i\Delta n\omega t} 
V^{(\pm)}(t),
\end{equation}
is cast into
\begin{equation}
\tilde{V}^{(\pm)}_{nn^\prime}
=
\left(
\begin{array}{cc}
0&
{1\over 2i}(\Omega_y\pm\Omega_x)\delta_{n,n^\prime+1}
\\
&-{1\over 2i}(\Omega_y\mp\Omega_x)\delta_{n,n^\prime-1}
\\
{1\over 2i}(\Omega_y\mp\Omega_x)\delta_{n,n^\prime+1}
\\
-{1\over 2i}(\Omega_y\pm\Omega_x)\delta_{n,n^\prime-1}
&0
\end{array}
\right).
\end{equation}

\section{Analytic expressions of energy dispersion $E(\boldsymbol{k})$}
\label{app2}
\subsection{Floquet Hamiltonian and approximate eigenvalues}
\label{app2-1}

The eigenvalue problem of the following Floquet Hamiltonian for up- and down-spins
\begin{equation}
L^{(\pm)}(\boldsymbol{k},t)=H^{(\pm)}(\boldsymbol{k},t)+V^{(\pm)}(t)-iI_2{\partial \over \partial t}
\label{L+-}
\end{equation}
is solved approximately to obtain analytic expressions of eigenvalue $E(\boldsymbol{k})$,
where
$H^{(\pm)}(\boldsymbol{k},t)$ and $V^{(\pm)}(t)$ are given in
Eqs.~(\ref{H+-}) and (\ref{V+-}), respectively.
The associated Fourier-Floquet matrix $\tilde{L}^{(\pm)}(\boldsymbol{k},\omega)$
is represented as
\begin{equation}
\tilde{L}^{(\pm)}(\boldsymbol{k},\omega)=
\left(
\begin{array}{@{\,}cccc@{\,}}
\ddots&\vdots&\vdots&\reflectbox{$\ddots$}\\
\ldots&\mathcal{L}^{(\pm)}_{n,n}&\mathcal{X}^{(\pm)}_{n,n-1}&\ldots\\
\ldots&\mathcal{X}^{(\pm)}_{n-1,n}&\mathcal{L}^{(\pm)}_{n-1,n-1}&\ldots\\
\reflectbox{$\ddots$}&\vdots&\vdots&\ddots
\end{array}
\right)
\end{equation}
where
$\mathcal{L}^{(\pm)}_{n,n}$ and $\mathcal{X}^{(\pm)}_{n,n-1}$ are $2\times 2$-block-matrices defined by
\begin{equation}
\mathcal{L}^{(\pm)}_{n,n}=
\left(
\begin{array}{@{\,}cccc@{\,}}
\tilde{L}^{(\pm)}_{1(n)1(n)}&\tilde{L}^{(\pm)}_{1(n)2(n+1)}\\
\tilde{L}^{(\pm)}_{2(n+1)1(n)}&\tilde{L}^{(\pm)}_{2(n+1)2(n+1)}
\end{array}
\right)
\label{mathcalL+-}
\end{equation}
and
\begin{equation}
\mathcal{X}^{(\pm)}_{n,n-1}=
\left(
\begin{array}{@{\,}cccc@{\,}}
\tilde{L}^{(\pm)}_{1(n)1(n-1)}&\tilde{L}^{(\pm)}_{1(n)2(n)}\\
\tilde{L}^{(\pm)}_{2(n+1)1(n-1)}&\tilde{L}^{(\pm)}_{2(n+1)2(n)}
\end{array}
\right)
\label{mathcalX+-}
\end{equation}
with
$
\mathcal{X}^{(\pm)}_{n-1,n}=[\mathcal{X}^{(\pm)}_{n,n-1}]^\dagger,
$
respectively.
Here, $\tilde{L}^{(\pm)}_{b(n)b^\prime(n^\prime)}$ is a Fourier-transform of
the $(b,b^\prime)$-matrix element of $L^{(\pm)}(\boldsymbol{k},t)$, that is,
\begin{equation}
\tilde{L}^{(\pm)}_{b(n)b^\prime(n^\prime)}
= {1\over T}\int_{0}^T dt \:e^{-i\Delta n\omega t} 
L^{(\pm)}_{bb^\prime}(\boldsymbol{k},t),
\end{equation}
where it is understood that the index $b=1(2)$ represents the conduction (valence) band.

Now, an approximation is made that $\tilde{L}^{(\pm)}(\boldsymbol{k},\omega)$ is replaced by
a sequence of $4\times 4$-block matrices $\{L^{(\pm)}_n\}$, 
\begin{equation}
\tilde{L}^{(\pm)}(\boldsymbol{k},\omega)\approx
\left(
\begin{array}{@{\,}ccccc@{\,}}
\ddots&&&&\\
&L^{(\pm)}_{n+1}&&O&\\
&&L^{(\pm)}_n&&\\
&^tO&&L^{(\pm)}_{n-1}&\\
&&&&\ddots
\end{array}
\right),
\label{tildeL+-}
\end{equation}
where
the off-diagonal elements are represented in terms of a null upper-triangular block $O$ and its transpose $^tO$, and the diagonal block matrix is given by
\begin{equation}
L_n^{(\pm)}=
\left(
\begin{array}{@{\,}cc@{\,}}
\mathcal{L}^{(\pm)}_{n,n}&\mathcal{X}^{(\pm)}_{n,n-1}\\
\mathcal{X}^{(\pm)}_{n-1,n}&\mathcal{L}^{(\pm)}_{n-1,n-1}
\end{array}
\right).
\end{equation}
It is obvious that this corresponds to the rotational-wave approximation, in which
just almost resonant coupling terms of $\mathcal{X}^{(\pm)}_{n,n-1}$ and $\mathcal{X}^{(\pm)}_{n-1,n}$ are sustained, and the rest terms are disregarded.
In Eq.~(\ref{mathcalL+-}), the diagonal elements of $\tilde{L}^{(\pm)}_{1(n)1(n)}$ and
$\tilde{L}^{(\pm)}_{2(n+1)2(n+1)}$ of $\mathcal{L}^{(\pm)}_{n,n}$ are strongly coupled by the off-diagonal
element, 
$\tilde{L}^{(\pm)}_{1(n)2(n+1)}$ and $\tilde{L}^{(\pm)}_{2(n+1)1(n)}$ representing interband interactions,
when an almost on-resonant condition is met.
Setting a set of eigenvalues and the associated eigenvectors of $\mathcal{L}^{(\pm)}_{n,n}$ 
as $\mathcal{E}^{(\pm)}_{n,\alpha}$ and $u^{(\pm)}_{n,\alpha}$ with $\alpha=1,2$, respectively, that is,
\begin{equation}
\mathcal{L}^{(\pm)}_{n,n}u^{(\pm)}_{n,\alpha}=\mathcal{E}^{(\pm)}_{n,\alpha}u^{(\pm)}_{n,\alpha},
\end{equation}
with
$U^{(\pm)}_n$ as a $2\times 2$-unitary matrix given by
\begin{equation}
U^{(\pm)}_n=\left(u^{(\pm)}_{n,1}u^{(\pm)}_{n,2}\right), 
\label{Upm+-}
\end{equation}
$\bar{L}^{(\pm)}_n$ defined as
\begin{equation}
\bar{L}^{(\pm)}_n=U^{(\pm)\dagger}_n L^{(\pm)}_nU^{(\pm)}_n
\end{equation}
becomes of the form
\begin{eqnarray}
\bar{L}^{(\pm)}_n
&=&
\left(
\begin{array}{@{\,}cc@{\,}}
\mathcal{E}^{(\pm)}_{n}&\bar{\mathcal{X}}^{(\pm)}_{n,n-1}\\
\bar{\mathcal{X}}^{(\pm)}_{n-1,n}&\mathcal{E}^{(\pm)}_{n-1}
\end{array}
\right) \nonumber\\
&=&
\left(
\begin{array}{@{\,}c|cc|c@{\,}}
\mathcal{E}^{(\pm)}_{n,1}&0&\times&\times\\\hline
0&\mathcal{E}^{(\pm)}_{n,2}&\mathcal{V}^{(\pm)}&\times\\
\times&\mathcal{V}^{(\pm)*}&\mathcal{E}^{(\pm)}_{n-1,1}&0\\\hline
\times&\times&0&\mathcal{E}^{(\pm)}_{n-1,2}
\end{array}
\right),
\label{barLn+-}
\end{eqnarray}
where
\begin{equation}
\mathcal{E}^{(\pm)}_n
=
\left(
\begin{array}{@{\,}cc@{\,}}
\mathcal{E}^{(\pm)}_{n,1}&0\\
0&\mathcal{E}^{(\pm)}_{n,2}
\end{array}
\right)
\end{equation}
and
\begin{equation}
\bar{\mathcal{X}}^{(\pm)}_{n.n-1}
=
U^{(\pm)\dagger}_n \mathcal{X}^{(\pm)}_{n,n-1}U^{(\pm)}_{n-1}
\equiv
\left(
\begin{array}{@{\,}cc@{\,}}
\times&\times\\
\mathcal{V}^{(\pm)}&\times
\end{array}
\right).
\label{mathcalV}
\end{equation}
In the second equality of the above equation,
$\mathcal{V}^{(\pm)}$ represents the (2,1)-components of $\bar{\mathcal{X}}^{(\pm)}_{n.n-1}$
and all other components are expressed just as a symbol $\times$;
$\mathcal{V}^{(\pm)}$ is independent of $n$, as shown later.
Here, it is supposed that just the component $\mathcal{V}^{(\pm)}$ is retained with neglecting the components denoted as $\times$.

Thus, one obtains two kinds of eigenenergies from Eq.~(\ref{barLn+-}), --- denoted as
$E^{(\pm)}_{n,\beta}(\boldsymbol{k})$ with $\beta=1,2$ ---, which are 
the eigenvalues of the $2\times 2$-block matrix 
\begin{equation}
\bar{L}^{(\pm)}_{D,n}\equiv
\left(
\begin{array}{@{\,}cc@{\,}}
\mathcal{E}^{(\pm)}_{n,2}&\mathcal{V}^{(\pm)}\\
\mathcal{V}^{(\pm)*}&\mathcal{E}^{(\pm)}_{n-1,1}
\end{array}
\right).
\label{LDn}
\end{equation}
Explicit expressions of $E^{(\pm)}_{n,\beta}$ are provided as
\begin{eqnarray}
E^{(\pm)}_{n,\beta}(\boldsymbol{k})
&=&
{1\over 2}\left[
\mathcal{E}^{(\pm)}_{n,2}+\mathcal{E}^{(\pm)}_{n-1,1}-(-1)^\beta
\right.\nonumber\\
&&\left.\times
\sqrt{
\left(\mathcal{E}^{(\pm)}_{n,2}-\mathcal{E}^{(\pm)}_{n-1,1}\right)^2+4|\mathcal{V}^{(\pm)}|^2
}
\right].
\label{En+-}
\end{eqnarray}
Therefore, the approximate eigenvalues of $\tilde{L}^{(\pm)}(\boldsymbol{k},\omega)$ of Eq.~(\ref{tildeL+-}) are given in terms of a series of the set of
eigenvalues 
$\{E^{(\pm)}_{n,1}(\boldsymbol{k}),E^{(\pm)}_{n,2}(\boldsymbol{k})\}$.

\subsection{Energy dispersions of FWSM states}
\label{app2-2}

The more detailed expressions of $E^{(\pm)}_{n,\beta}(\boldsymbol{k})$ are sought.
To do this, $\mathcal{E}^{(\pm)}_{n,\alpha}$ is represented in terms of 
the matrix elements of Eq.~(\ref{mathcalL+-}) as
\begin{equation}
\mathcal{E}^{(\pm)}_{n,\alpha}
=
{1\over 2}\left[
\epsilon^{0(\pm)}_n-(-1)^\alpha
\sqrt{
\left( \Delta\epsilon^{(\pm)}_n \right)^2+4|\mathcal{W}^{(\pm)}|^2
}
\right],
\label{mathcalEn+-}
\end{equation}
where
\begin{equation}
\epsilon^{0(\pm)}_n=\tilde{L}^{(\pm)}_{1(n)1(n)}+\tilde{L}^{(\pm)}_{2(n+1)2(n+1)},
\end{equation}
\begin{equation}
\Delta\epsilon^{(\pm)}_n=\tilde{L}^{(\pm)}_{1(n)1(n)}-\tilde{L}^{(\pm)}_{2(n+1)2(n+1)},
\end{equation}
and
\begin{equation}
\mathcal{W}^{(\pm)}=\tilde{L}^{(\pm)}_{1(n)2(n+1)}.
\end{equation}
According to the explicit expressions of 
$\tilde{D}_{j,nn^\prime}(\boldsymbol{k},\omega)$
obtained in Appendix~\ref{app1}, it is shown that
\begin{equation}
\epsilon^{0(\pm)}_n=(2n+1)\omega
\end{equation}
and
\begin{equation}
\Delta\epsilon^{(\pm)}_n\equiv 2\eta(\boldsymbol{k})=2D(\boldsymbol{k})-\omega,
\end{equation}
where
\begin{eqnarray}
D(\boldsymbol{k})
&=&
\Delta_g+\epsilon_z(k_z)
\nonumber\\
&&
-\sum_{l=x,y}2t^{xy}\left[1-J_0(z_l)\cos{(k_ld_l)} \right],
\end{eqnarray}
with
\begin{equation}
\epsilon_z(k_z)=
-2t^z\left[1-\cos{(k_zd_z)} \right].
\label{epsilonz}
\end{equation}
Similarly, in view of $\tilde{V}^{(\pm)}_{nn^\prime}$ given in Appendix~\ref{app1}, one has
\begin{eqnarray}
\left\{
\begin{array}{l}
\mathcal{W}^{(+)}=-{i\over 2}\Omega_x-it_{sp}J_1(z_x)\cos{(k_xd_x)}\\
\mathcal{W}^{(-)}=(\mathcal{W}^{(+)})^*
\end{array}
\right.
\label{mathcalWl}
\end{eqnarray}
for the linearly polarized light,
and
\begin{eqnarray}
\left\{
\begin{array}{lll}
\mathcal{W}^{(+)}&=&-it_{sp}J_1(z_c)\left[\cos{(k_xd_x)}+\cos{(k_yd_y)}\right]\\
\mathcal{W}^{(-)}&=&it_{sp}J_1(z_c)\left[\cos{(k_xd_x)}-\cos{(k_yd_y)}\right]\\
&&+i\Omega_c
\end{array}
\right.
\label{mathcalWc}
\end{eqnarray}
for the circularly polarized light,
with $\Omega_c\equiv\Omega_x=\Omega_y$.

Further, $U^{(\pm)}_n$ of Eq.~(\ref{Upm+-}) is given by
\begin{equation}
u^{(\pm)}_{n,1}
=
\left(
\begin{array}{l}
\cos{\Theta^{(\pm)}}\\
\sin{\Theta^{(\pm)}}e^{-i\Phi^{(\pm)}}
\end{array}
\right)
\end{equation}
and
\begin{equation}
u^{(\pm)}_{n,2}
=
\left(
\begin{array}{l}
\sin{\Theta^{\prime(\pm)}}\:e^{i\Phi^{(\pm)}}\\
\cos{\Theta^{\prime(\pm)}}
\end{array}
\right),
\end{equation}
where
\begin{equation}
\tan{\Theta^{(\pm)}}=
{|\mathcal{W}^{(\pm)}|\over
\eta(\boldsymbol{k})+
\sqrt{
\eta(\boldsymbol{k})^2+|\mathcal{W}^{(\pm)}|^2
}
},
\end{equation}
\begin{equation}
\tan{\Theta^{\prime(\pm)}}=-\tan{\Theta^{(\pm)}},
\end{equation}
and
\begin{equation}
e^{i\Phi^{(\pm)}}={\mathcal{W}^{(\pm)}\over |\mathcal{W}^{(\pm)}|}.
\end{equation}
Thus, $\mathcal{V}^{(\pm)}$ of Eq.~(\ref{mathcalV}) is approximately given by
\begin{equation}
\mathcal{V}^{(\pm)}
\approx
\tilde{L}^{(\pm)}_{1(n)2(n)}\sin{\Theta^{(\pm)}}\sin{\Theta^{\prime(\pm)}}e^{-2i\Phi^{(\pm)}},
\label{mathcalV2}
\end{equation}
where just
the most dominant component
$\tilde{L}^{(\pm)}_{1(n)2(n)}$
in $\mathcal{X}^{(\pm)}_{n,n-1}$ of Eq.~(\ref{mathcalX+-})
is kept under the condition that
$J_0(z_l) \gg J_{n(\ge1)}(z_l)$ in the range of $z_l$ concerned here.
To be more specific,
$\tilde{L}^{(\pm)}_{1(n)2(n)}$ is represented as
\begin{equation}
\tilde{L}^{(\pm)}_{1(n)2(n)}=\pm t_{sp}\left[J_0(z_x)\sin{(k_xd_x)}\mp i\sin{(k_yd_y)}\right]
\end{equation}
for the the linearly polarized light,
and
\begin{equation}
\tilde{L}^{(\pm)}_{1(n)2(n)}=\pm t_{sp}J_0(z_c)\left[\sin{(k_xd_x)}\mp i\sin{(k_yd_y)}\right]
\end{equation}
for the the circularly polarized light.

According to the above results, 
$E^{(\pm)}_{n,\beta}(\boldsymbol{k})$
of Eq.~(\ref{En+-}) is cast into
\begin{equation}
E^{(\pm)}_{n,\beta}(\boldsymbol{k})=n\omega+E^{(\pm)}_{0,\beta}(\boldsymbol{k}),
\label{Ebeta}
\end{equation}
where
\begin{eqnarray}
E^{(\pm)}_{0,\beta}(\boldsymbol{k})
&=&(-1)^{\beta-1}
\left[\left\{
\omega/2
-\left(\eta(\boldsymbol{k})^2+|\mathcal{W}^{(\pm)}|^2
\right)^{1/2}
\right\}^2
\right.\nonumber\\
&&
\left.
+|\mathcal{V}^{(\pm)}|^2
\right]^{1/2}.
\label{E0beta}
\end{eqnarray}
Obviously, it is likely that these two photon sidebands, $E^{(\pm)}_{n,1}(\boldsymbol{k})$ and $E^{(\pm)}_{n^\prime,2}(\boldsymbol{k})$, touch each other 
when the following conditions are met:
$n=n^\prime$ and the expression inside the square brackets of Eq.~(\ref{E0beta}) vanishes.
The second condition is ensured only when $k_x=k_y=0$, that is, $\mathcal{V}^{(\pm)}=0$.
Thus, this becomes the existence condition of a touching point at
$k_z^{D/W(\pm)}$ satisfying the relation
\begin{equation}
\epsilon_z(k_z^{D/W(\pm)})=\omega/2-\Delta^\prime_g+\sqrt{(\omega/2)^2-|\mathcal{W}_0^{(\pm)}|^2},
\label{node}
\end{equation}
where
\begin{equation}
\Delta^\prime_g=\Delta_g-\sum_{l=x,y}2t^{xy}[1-J_0(z_l)]
\end{equation}
and 
\begin{equation}
\mathcal{W}_0^{(\pm)}=\mathcal{W}^{(\pm)}\left|_{k_x=k_y=0}\right..
\end{equation}
Here, $k_z^{D(\pm)}$ represents a solution of Eq.~(\ref{node}) for the drive of
linearly polarized laser, while $k_z^{W(\pm)}$ for the drive of
circularly polarized laser.
In view of Eq.~(\ref{epsilonz}), a certain pair of values, $k_z^{D/W(\pm)}$ and 
$-k_z^{D/W(\pm)}$, exists under the condition that
\begin{eqnarray}
-\omega/2+\Delta^\prime_g&<&\sqrt{(\omega/2)^2-|\mathcal{W}_0^{(\pm)}|^2}
\nonumber\\
&<&-\omega/2+\Delta^\prime_g-4t^z
\label{nodecond0}
\end{eqnarray}
with $t^z< 0$,
and
\begin{equation}
\omega/2 >|\mathcal{W}_0^{(\pm)}|.
\end{equation}
The above condition is recast into
\begin{eqnarray}
&\text{Max}&\left(
2|\mathcal{W}_0^{(\pm)}|,
\frac{\Delta^{\prime2}_g+|\mathcal{W}_0^{(\pm)}|^2}
{\Delta^{\prime}_g}
\right)
<\omega
\nonumber\\
&&<
\frac{(\Delta^{\prime}_g-4t^z)^2+|\mathcal{W}_0^{(\pm)}|^2}
{\Delta^{\prime}_g-4t^z}.
\label{nodecond}
\end{eqnarray}
It is evident that these pairs of values $k_z^{D/W(\pm)}$ and $-k_z^{D/W(\pm)}$
correspond to nodal points lying on the $\Gamma-Z$ axis in the BZ
for FDSM/FWSM states.

\subsection{Band structures of $E^{(\pm)}_{hh(1)}(\boldsymbol{k},\omega)$ and $E^{(\pm)}_{e(-1)}(\boldsymbol{k},\omega)$ }
\label{app2-3}

Here, the label $b(n)$ (with $b=e,hh$) is introduced to represent a Floquet state attributed to $b$-band with $n$ photon dressing;
$e$ and $hh$ mean electron and heavy-hole bands, respectively.
Following this, a Floquet energy $E^{(\pm)}(\boldsymbol{k})$ is given by
\begin{equation}
E^{(\pm)}_{hh(n+1)}(\boldsymbol{k})\equiv E^{(\pm)}_{n,1}(\boldsymbol{k})\ge n\omega
\label{Ehhn+1}
\end{equation}
and
\begin{equation}
E^{(\pm)}_{e(n-1)}(\boldsymbol{k})\equiv E^{(\pm)}_{n,2}(\boldsymbol{k})\le n\omega
\label{Een-1}
\end{equation}
in the case that 
a Floquet band $hh(n+1)$ is located above a Floquet band $e(n-1)$.
In the opposite case that $hh(n+1)$ is located below $e(n-1)$, it is understood that
the above definition of $ E^{(\pm)}_{hh(n+1)}(\boldsymbol{k})$ is replaced by
that of $ E^{(\pm)}_{e(n-1)}(\boldsymbol{k})$.

Below, an energy dispersion of $E^{(\pm)}(\boldsymbol{k})=E^{(\pm)}_{hh(1)}(\boldsymbol{k})\ge0$ in the vicinity of $\boldsymbol{k}^{D/W(\pm)}
\equiv (0,0,k_z^{D/W(\pm)})$ is examined, where
$E^{(\pm)}_{e(-1)}(\boldsymbol{k})=-E^{(\pm)}_{hh(1)}(\boldsymbol{k})$.
To do this, $\eta(\boldsymbol{k})$, $\mathcal{W}^{(\pm)}$, and
$\mathcal{V}^{(\pm)}$ are expanded around this point as follows:
\begin{eqnarray}
\eta(\boldsymbol{k})
&\approx& \eta_0+\eta_z^{(1)}(\Delta k_zd_z)\nonumber\\
&&+\sum_{l=x,y}\eta_l(k_ld_l)^2+\eta_z(\Delta k_zd_z)^2,
\label{etaexp}
\end{eqnarray}
\begin{equation}
\mathcal{W}^{(\pm)}
\approx \mathcal{W}^{(\pm)}_0+\sum_{l=x,y}\omega^{(\pm)}_l(k_ld_l)^2,
\label{Wexp}
\end{equation}
and
\begin{equation}
\mathcal{V}^{(\pm)}
\approx \sum_{l=x,y}\nu^{(\pm)}_l(k_ld_l),
\label{Vexp}
\end{equation}
where $\Delta k_z=k_z-k_z^{D/W(\pm)}$, and
the expansion coefficients $\eta_0, \eta_z^{(1)}, \eta_l, \eta_z, \mathcal{W}^{(\pm)}_0, \omega_l^{(\pm)}$, and $\nu_l^{(\pm)}$ are represented by
\begin{eqnarray}
\eta_0&=&\Delta^\prime_g+\epsilon_z(k_z^{D/W(\pm)})-\omega/2,
\eta_z^{(1)}=-2t^z\sin{(k_z^{D/W(\pm)}d_z)},\nonumber\\
\eta_l&\approx&-t^{xy}, \eta_z=-t^z\cos{(k_z^{D/W(\pm)}d_z)},
\label{eta}
\end{eqnarray}
\begin{eqnarray}
\left\{
\begin{array}{lll}
\mathcal{W}^{(+)}_0=-{i\over 2}\Omega_x,&
\mathcal{W}^{(-)}_0={i\over 2}\Omega_x&\text{(linear)}\\
\mathcal{W}^{(+)}_0=0,&\mathcal{W}^{(-)}_0=i\Omega_c&\text{(circular)}
\end{array}
\right.,
\label{W0+-}
\end{eqnarray}
\begin{equation}
\omega_l^{(\pm)}\approx 0
\label{omegal}
\end{equation}
and
\begin{equation}
\nu_x^{(\pm)}=\pm\sigma^{(\pm)}t_{sp},\;\;
\nu_y^{(\pm)}=-i\sigma^{(\pm)}t_{sp},
\label{nul}
\end{equation}
respectively.
It is considered that $J_0(z_l)\approx 1$ in the range of $z_l$ concerned here
in Eq.~(\ref{eta}), 
terms including $t_{sp}J_1(z_l)$ are neglected
due to $|J_1(z_l)|\ll 1$ in Eqs.~(\ref{W0+-}) and (\ref{omegal}), and
the prefactor $\sigma^{(\pm)}$ is given by
\begin{equation}
\sigma^{(\pm)}=\sin{\Theta^{(\pm)}_0}\sin{\Theta^{\prime(\pm)}_0}e^{-2i\Phi^{(\pm)}_0}
\label{sigma+-}
\end{equation}
in Eq.~(\ref{nul}),
where
\begin{eqnarray}
&&\Theta^{(\pm)}_0=\Theta^{(\pm)}|_{\boldsymbol{k}=\boldsymbol{k}^{D/W(\pm)}},
\;\;
\Theta^{\prime(\pm)}_0=\Theta^{\prime(\pm)}|_{\boldsymbol{k}=\boldsymbol{k}^{D/W(\pm)}},
\nonumber\\
&&
\Phi^{(\pm)}_0=\Phi^{(\pm)}|_{\boldsymbol{k}=\boldsymbol{k}^{D/W(\pm)}}.
\end{eqnarray}

For the FDSM driven by the linearly polarized light,
\begin{eqnarray}
&&E^{(\pm)}(\boldsymbol{k})\nonumber\\
&&\approx
\left[
\left|
\sum_{l=x,y}\nu_l^{(\pm)}(k_ld_l)
\right|^2
+\frac{(\eta_0\eta_z^{(1)})^2}
{(\omega/2)^2}(\Delta k_zd_z)^2
\right]^{1/2}\nonumber\\
&&=
\left[
\sum_{l=x,y}(\xi^D_l)^2(k_ld_l)^2+(\xi^D_z)^2(\Delta k_zd_z)^2
\right]^{1/2},
\label{Dcone}
\end{eqnarray}
where
$\Delta k_z=k_z-k_z^{D(\pm)}$, 
$\xi^D_x=\xi^D_y=|\sigma^{(\pm)}|t_{sp}$ with
\begin{eqnarray}
|\sigma^{(\pm)}|=\frac{(\Omega_x/2)^2}
{
(\Omega_x/2)^2
+\left(\eta_0
+\sqrt{\eta_0^2+(\Omega_x/2)^2}
\right)^2
},
\end{eqnarray}
and $\xi^D_z=|\eta_0\eta_z^{(1)}|/(\omega/2)$.
Hence, it is seen that the Floquet bands of $E^{(\pm)}_{hh(1)}(\boldsymbol{k})$ and
$E^{(\pm)}_{e(-1)}(\boldsymbol{k})$ cross linearly 
at the Dirac points of $\boldsymbol{k}^{D(\pm)}$ and $-\boldsymbol{k}^{D(\pm)}$.
On the other hand,
for the FWSM driven by the circularly polarized light,
\begin{eqnarray}
&&E^{(+)}(\boldsymbol{k})\nonumber\\
&&\approx
\left|
\sum_{l=x,y}\eta_l(k_ld_l)^2+\eta_z^{(1)}(\Delta k_zd_z)+\eta_z(\Delta k_zd_z)^2
\right|\nonumber\\
&&=
\left|
(-t^{xy})\sum_{l=x,y}(k_ld_l)^2
+\eta_z^{(1)}(\Delta k_zd_z)
+\eta_z(\Delta k_zd_z)^2
\right|,\nonumber
\\
\label{W+cone}
\end{eqnarray}
due to $\mathcal{W}^{(+)}_0=0$ and
$\nu_x^{(+)}=i\nu_y{(+)}=\sigma^{(+)}t_{sp}\approx 0$
with $|\sigma^{(+)}|=[t_{sp}J_1(z_c)/(2\eta_0)]^2$,
where $\Delta k_z=k_z-k_z^{W(+)}$, 
while
\begin{eqnarray}
&&E^{(-)}(\boldsymbol{k})\nonumber\\
&&\approx
\left[
\left|
\sum_{l=x,y}\nu_l^{(-)}(k_ld_l)
\right|^2
+\frac{(\eta_0\eta_z^{(1)})^2}
{(\omega/2)^2}(\Delta k_zd_z)^2
\right]^{1/2}\nonumber\\
&&=
\left[
\sum_{l=x,y}(\xi^W_l)^2(k_ld_l)^2+(\xi^W_z)^2(\Delta k_zd_z)^2
\right]^{1/2},
\label{W-cone}
\end{eqnarray}
where
$\Delta k_z=k_z-k_z^{W(-)}$, 
$\xi^W_x=\xi^W_y=|\sigma^{(-)}|t_{sp}$ with
\begin{eqnarray}
|\sigma^{(-)}|=\frac{(\Omega_c)^2}
{
(\Omega_c)^2
+\left(\eta_0
+\sqrt{\eta_0^2+(\Omega_c)^2}
\right)^2
},
\end{eqnarray}
and $\xi^W_z=|\eta_0\eta_z^{(1)}|/(\omega/2)$.
Hence, it is seen that 
the Floquet bands of $E^{(+)}_{hh(1)}(\boldsymbol{k})$ and
$E^{(+)}_{e(-1)}(\boldsymbol{k})$ cross in a quadratic manner
in the $k_x$- and $k_y$-directions and in a linear manner in the $k_z$-direction
at the Weyl points of $\boldsymbol{k}^{W(+)}$ and $-\boldsymbol{k}^{W(+)}$.
On the other hand,
those of $E^{(-)}_{hh(1)}(\boldsymbol{k})$ and
$E^{(-)}_{e(-1)}(\boldsymbol{k})$ cross in a linear manner
at the Weyl points of $\boldsymbol{k}^{W(-)}$ and $-\boldsymbol{k}^{W(-)}$,
similar to the FDSMs.

Finally, band structures in the vicinity of the $k_x-k_y$ plane 
with $k_z$ fixed are examined.
For the FWSM of the up-spin state, 
$E^{(+)}(\boldsymbol{k})$ is represented simply as 
\begin{equation}
E^{(+)}(\boldsymbol{k})
\approx
|\omega/2-D(\boldsymbol{k})|
\end{equation}
due to Eq.~(\ref{E0beta}),
where
$\mathcal{W}^{(+)}$ and $\mathcal{V}^{(+)}$ are neglected within the order of $t_{sp}J_1(z_c)\approx 0$.
Thus, $E^{(+)}_{hh(1)}(\boldsymbol{k})\approx 0$ for $\boldsymbol{k}$'s situated 
on the closed surface
\begin{eqnarray}
&&-2t^{xy}\sum_{l=x,y}[1-J_0(z_l)\cos{(k_ld_l)}]
-2t^z[1-\cos{(k_zd_z)}]\nonumber\\
&&=\omega-\Delta_g >0.
\label{closedsurf}
\end{eqnarray}
This shows that there exists a nodal ring in the $k_x-k_y$ plane, which is reminiscent of a NLSM phase in the FWSM of concern.
As regards the FWSM of the down-spin state,
$E^{(-)}_{hh(1)}(\boldsymbol{k})$ and
$E^{(-)}_{e(-1)}(\boldsymbol{k})$ are gapped out
in the $k_x-k_y$ plane due to 
$\mathcal{V}^{(-)}$ that is not negligible.
It is remarked that an accidental band crossing occurs between these two bands 
at high-symmenty points of $\boldsymbol{k}^*=(0,0,0), (0,\pm\pi/d_y,0), (\pm\pi/d_x,0,0)$, and $(\pm\pi/d_x,\pm\pi/d_y,0)$ at which $\mathcal{V}^{(-)}=0$,
only when the relation
\begin{equation}
\omega/2=\sqrt{\eta(\boldsymbol{k}^*)^2+\Omega_c^2}
\end{equation}
is ensured.
In passing, the similar result with that of the FWSM of the down-spin state is applied for 
the FDSMs.

\subsection{Floquet-Weyl Hamiltonian and Chirality}
\label{app2-4}

Here, effective Floquet-Weyl Hamiltonians the eigenvalues of which are given by
Eqs.~(\ref{W+cone}) and (\ref{W-cone}) are derived from Eq.~(\ref{LDn}) in the vicinity of
the Weyl points $\boldsymbol{k}^{W(+)}$ and $\boldsymbol{k}^{W(-)}$.
The Hamiltonian for the up-spin band, represented as 
$L_n^{W(+)}\equiv \bar{L}_{D,n}^{(+)}$, is cast into
\begin{eqnarray}
L_n^{W(+)}
&=&n\omega+\left(
\begin{array}{cc}
-\eta_z^{(1)}(\Delta k_zd_z)&\sum_{l=x,y}\nu^{(+)}_l(k_ld_l)\\
\sum_{l=x,y}\nu_l^{(+)*}(k_ld_l)&\eta_z^{(1)}(\Delta k_zd_z)
\end{array}
\right)
\nonumber\\
&=&n\omega+h_+^{(+)}\boldsymbol{p}^{(+)}\cdot\boldsymbol{\sigma},
\label{LW+}
\end{eqnarray}
where Eqs.~(\ref{mathcalEn+-}), (\ref{etaexp}), (\ref{Wexp}), and (\ref{Vexp}) are used.
Further, the effective momentum $\boldsymbol{p}^{(+)}$, defined in the right-handed system,
is given by
\begin{eqnarray}
p_l^{(+)}&=&h_+^{(+)}\sum_{l^\prime=x,y,z}v_{ll^\prime}^{(+)}(k_{l^\prime}-k_{l^\prime}^{W(+)})
\nonumber\\
&\equiv&h_+^{(+)}\boldsymbol{v}_{l}^{(+)}\cdot(\boldsymbol{k}-\boldsymbol{k}^{W(+)}),
\end{eqnarray}
where
\begin{eqnarray}
\left\{
\begin{array}{l}
\boldsymbol{v}_x^{(+)}=(\nu^{(+)}_xd_x,0,0)\\
\boldsymbol{v}_y^{(+)}=(0,\nu^{(+)}_xd_y,0)\\
\boldsymbol{v}_z^{(+)}=(0,0,-\eta_z^{(1)}d_z)
\end{array}
\right.
\end{eqnarray}
and
$h_+^{(+)}$ is a helicity of particle at
the Weyl point $\boldsymbol{k}^{W(+)}$,
defined by
\begin{equation}
h_+^{(+)}=\text{sgn}(\boldsymbol{v}_x^{(+)}\times\boldsymbol{v}_y^{(+)}
\cdot\boldsymbol{v}_z^{(+)}),
\label{h++}
\end{equation}
which is either +1 or -1.
Thus, the positive eigenvalue of Eq.~(\ref{LW+}) for $n=0$ is given by
\begin{eqnarray}
E^{W(+)}(\boldsymbol{k})&=&
\left|
(\eta_z^{(1)})^2(\Delta k_zd_z)^2+\nu_x^{(+)2}\sum_{l=x,y}(k_ld_l)^2
\right|^{1/2}\nonumber\\
&\approx&\left|
\eta_z^{(1)}(\Delta k_zd_z)
\right|,
\end{eqnarray}
where in the second equality the fact that $\nu^{(+)}_x \approx 0$ is considered.
This is identical to Eq.~(\ref{W+cone}) within the first order with respect to
$k_x, k_y,$ and $\Delta k_z$.

Similarly,
the Hamiltonian for the down-spin band, represented as 
$L_n^{W(-)}\equiv \bar{L}_{D,n}^{(-)}$, is cast into
\begin{eqnarray}
L_n^{W(-)}
&=&n\omega+\left(
\begin{array}{cc}
-{\eta_0\eta_z^{(1)}\over(\omega/2)}(\Delta k_zd_z)&\sum_{l=x,y}\nu^{(-)}_l(k_ld_l)\\
\sum_{l=x,y}\nu^{(-)*}_l(k_ld_l)&{\eta_0\eta_z^{(1)}\over(\omega/2)}(\Delta k_zd_z)
\end{array}
\right)
\nonumber\\
&=&n\omega+h_+^{(-)}\boldsymbol{p}^{(-)}\cdot\boldsymbol{\sigma}.
\label{LW-}
\end{eqnarray}
Here, the effective momentum $\boldsymbol{p}^{(-)}$, defined in the right-handed system,
is given by
\begin{eqnarray}
p_l^{(-)}&=&h_+^{(-)}\sum_{l^\prime=x,y,z}v_{ll^\prime}^{(-)}(k_{l^\prime}-k_{l^\prime}^{W(-)})
\nonumber\\
&\equiv&h_+^{(-)}\boldsymbol{v}_{l}^{(-)}\cdot(\boldsymbol{k}-\boldsymbol{k}^{W(-)}),
\end{eqnarray}
where
\begin{eqnarray}
\left\{
\begin{array}{l}
\boldsymbol{v}_x^{(-)}=(\nu^{(-)}_xd_x,0,0)\\
\boldsymbol{v}_y^{(-)}=(0,-\nu^{(-)}_xd_y,0)\\
\boldsymbol{v}_z^{(-)}=(0,0,-{\eta_0\eta_z^{(1)}\over(\omega/2)}d_z)
\end{array}
\right.
\end{eqnarray}
and
$h_+^{(-)}$ is a helicity of particle at
the Weyl point $\boldsymbol{k}^{W(-)}$, defined by
\begin{equation}
h_+^{(-)}=\text{sgn}(\boldsymbol{v}_x^{(-)}\times\boldsymbol{v}_y^{(-)}
\cdot\boldsymbol{v}_z^{(-)}),
\label{h+-}
\end{equation}
which is either +1 or -1.
It is obvious that the positive eigenvalue of Eq.~(\ref{LW-}) for $n=0$,
$E^{W(-)}(\boldsymbol{k})$, 
is identical to Eq.~(\ref{W-cone}).

Finally, the chiralities of FWSMs for both up- and down-spins are examined.
Let the helicities of the Weyl cones at $-\boldsymbol{k}^{W(+)}$ and $-\boldsymbol{k}^{W(-)}$
be represented as $h_-^{(+)}$ and $h_-^{(-)}$, respectively.
It is evident that $h_+^{(+)}h_-^{(+)}=-1$ and $h_+^{(-)}h_-^{(-)}=-1$,
since according to Eqs~(\ref{eta}) and (\ref{nul}), the replacement of 
the nodal position
at $\boldsymbol{k}^{W(\pm)}$ by that at $-\boldsymbol{k}^{W(\pm)}$ still keeps 
$\nu_x^{(\pm)}$ unaltered, whereas $\eta_z^{(1)}$ changes its sign; a double sign corresponds.
Further, it is also seen that $h_+^{(+)}h_+^{(-)}=-1$ and $h_-^{(+)}h_-^{(-)}=-1$,
since the sign of $\eta_z^{(1)}$ remains unaltered, whereas 
the sign of $\nu_{\pm}^{(+)}$ is different from that of $\nu_{\pm}^{(-)}$;
a double sign corresponds.
Therefore, it is verified that each of four pairs of the Weyl cones at $\pm\boldsymbol{k}^{W(+)}$,
$\pm\boldsymbol{k}^{W(-)}$, $\boldsymbol{k}^{W(\pm)}$, and $-\boldsymbol{k}^{W(\pm)}$
possesses opposite chiralities.

\section{Enlarged view of Figs.~\ref{fig3}(a) and \ref{fig3}(b)}
\label{app3}

Energy dispersions shown in Figs.~\ref{fig3}(a) and \ref{fig3}(b)} are enlarged 
in Fig.~\ref{fig7}
to make clearer the difference of band gaps between 
the up-spin and down-spin bands around
$E=0$.

\begin{figure}[t]
\begin{center}
\includegraphics[width=8cm,clip]{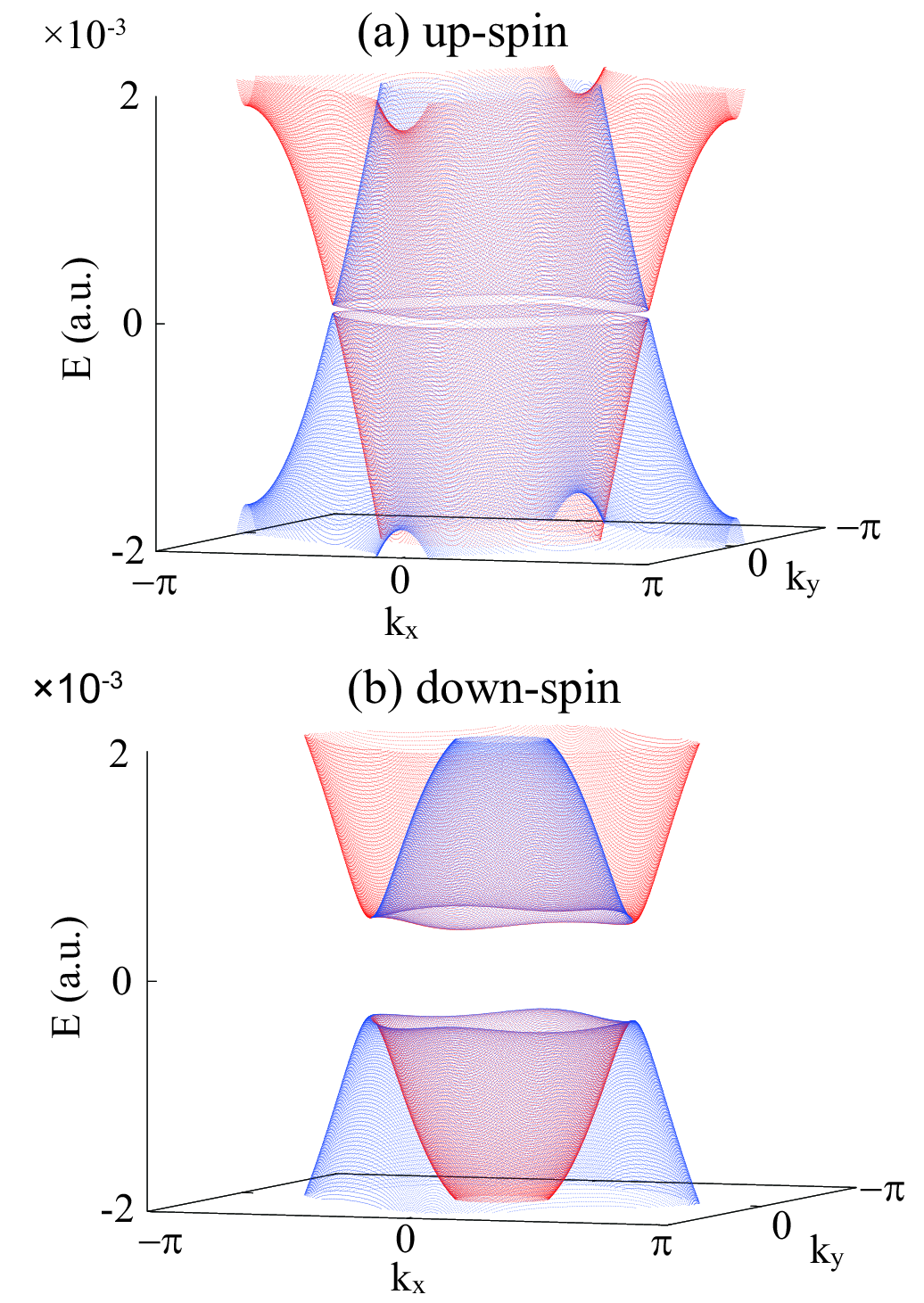}
\caption{Energy dispersions $E(\boldsymbol{k})$ of up-spin and down-spin bands 
at $k_z=0$ which are enlarged around $E=0$.
Here, bands dominated rather by the $s/p$-orbital component are denoted 
by a red/blue solid line.
The axis of abscissa $k_l$ is gauged in the unit of $1/d_l$ with $l=x,y,z$.
(a) $E(\boldsymbol{k})$ in the $k_x-k_y$ plane at $k_z=0$ for the up-spin bands.
(b) The same as panel (a) but for the down-spin bands.
}
\label{fig7}
\end{center}
\end{figure}



\end{document}